\documentclass[aps,prd,twocolumn,superscriptaddress,nofootinbib,10pt,floatfix]{revtex4-2}

\usepackage{amsmath,amssymb,amsfonts}
\usepackage{graphicx}
\usepackage{bm}
\usepackage{xcolor}
\usepackage[colorlinks=true,allcolors=blue]{hyperref}
\hypersetup{pdftitle={Periodic Dissipation in Warm Inflation: Coupled Response, Log-Periodic Gravitational Waves and Primordial Black Holes},pdfauthor={Mayukh R. Gangopadhyay}}

\makeatletter
\let\@bibdataout@init\relax
\let\auto@bib\@empty
\let\auto@bib@innerbib\@empty
\let\write@bibliographystyle\relax
\makeatother
\newcommand{\Pz}{\mathcal{P}_\zeta}
\newcommand{\Mpc}{\,\mathrm{Mpc}}
\newcommand{\Msun}{M_\odot}
\newcommand{\kpk}{k_{\rm pk}}
\newcommand{\fpbh}{f_{\rm PBH}}
\newcommand{\Ogw}{\Omega_{\rm GW}}
\newcommand{\order}[1]{\mathcal{O}(#1)}

\begin{document}

\title{Signatures of Periodic Dissipation in Warm Inflation: Power Spectrum, Non-Gaussianity, Gravitational Waves and Primordial Black Holes.}

\author{Mayukh R. Gangopadhyay}
\email{mayukhraj.g@vit.ac.in}
\affiliation{Department of Physics, School of Advanced Sciences, Vellore Institute of Technology (VIT), Chennai Campus, Chennai 600127, Tamil Nadu, India}

\date{\today}

\begin{abstract}
In warm inflation the inflaton loses energy to a hot bath of radiation, and this loss acts as a friction on its motion. If the couplings behind the friction depend periodically on the inflaton, as an approximate shift symmetry allows, the friction rises and falls as the field rolls. We study what this does to the primordial fluctuations, following the coupled fluctuations of the inflaton, the radiation and the metric on a plateau potential. The scales seen in the cosmic microwave background are unaffected, because the friction is negligible when they leave the horizon, and the model gives $n_s\simeq0.964$ and $r\simeq5\times10^{-3}$; reshaping the plateau there raises $n_s$ to $0.974$ without changing the rest. On smaller scales, where the friction stays strong for several e-folds, its oscillation leaves a regular, log-periodic ripple on the power spectrum, a comb that changes the power by about 34 percent. The size of the comb depends on how the coupled system responds to a friction that varies in time. We measure this response directly and find that it predicts the comb to within 7 percent across five versions of the model. The comb also appears in the three-point function, which oscillates with it, and in the gravitational waves induced by the enhanced fluctuations between $0.2$~Hz and $0.9$~kHz, where an idealized detector of the BBO class could resolve it with a signal-to-noise ratio of 6--16 if the foreground from neutron-star binaries is removed. A separate, localized rise of the friction gives a peak in the spectrum that, for a tuned height, produces asteroid-mass primordial black holes making up a large share of the dark matter, together with induced gravitational waves near $19$~mHz.
\end{abstract}

\maketitle

\section{Introduction}
\label{sec:intro}

Warm inflation drops the assumption that the inflaton is decoupled from everything else while it drives the expansion~\cite{BereraFang1995,Berera1995,BereraGleiserRamos1998}. If the inflaton dissipates even weakly into other fields, a subdominant radiation bath survives against the expansion, and a friction coefficient $\Upsilon$ adds to the Hubble drag~\cite{BereraMossRamos2009,BasteroGilBerera2009,Berera1997}. Whether such a bath can be sustained was contested early~\cite{YokoyamaLinde1999} and later settled by field-theory constructions that compute $\Upsilon$ instead of postulating it~\cite{BGRR2011diss,BereraRamos2005,BereraKephart1999,LaineProcacci2021}, such as the Warm Little Inflaton~\cite{WLI2016,BGRR2013,ReliableEFT2021}. The scenario now meets CMB data on the same footing as cold inflation~\cite{BenettiRamos2017,BasteroGilCMB2018,Planck2018X,BereraMabillard2018}, and it has been carried into quintessential inflation and baryogenesis~\cite{GangopadhyayWQI2021,BasakWQI2022}.

For an axion-like inflaton one ingredient has gone largely unused. The shift symmetry $\phi\to\phi+2\pi f$ that flattens the potential~\cite{FreeseFriemanOlinto1990} also shapes the couplings to the bath, and if a mass or a coupling in the dissipating sector depends on $\phi/f$, the friction inherits that periodicity. In cold inflation the analogous periodicity of the potential is well studied: it makes the spectrum oscillate and generates a resonant, log-periodic bispectrum~\cite{SilversteinWestphal2008,McAllister2010,FlaugerOscillations2010,FlaugerPajer2011,Hannestad2010,FlaugerDrifting2017,Behbahani2012}, whose shapes and detectability have been mapped in detail~\cite{ChenEastherLim2008,ChenReview2010,KobayashiTakahashi2011,Achucarro2011,ChenFeatures2012,PajerPeloso2013} against the single-field bispectrum and its shape classification~\cite{Maldacena2003,AcquavivaBartolo2003,BabichCreminelli2004,ChenHuangKachruShiu2007}. Warm inflation, including warm natural inflation~\cite{WarmNatural2023}, has been studied almost only with a smooth $\Upsilon$. The closest relatives are a Warm Little Inflaton whose fermion mass oscillates with the inflaton, which makes the slow-roll parameter $\eta$ oscillate while leaving the CMB scales untouched~\cite{FerrazRosa2023}, and preheated inflation, in which scalars with an oscillating mass are produced non-thermally by a narrow parametric resonance~\cite{GorgulhoRosa2025}. In the second case the produced particles do not form a thermal bath and do not act as a Markovian friction. Here we take a periodic friction at face value and follow it into the small-scale spectrum, the gravitational waves and the bispectrum.

Warm inflation is known to enhance small-scale power, to form primordial black holes and to source induced gravitational waves~\cite{Arya2020,BasteroGilSubias2021,CorreaWNI2022,Correa2024GW,AryaJainMishra2024,FerrazMansoRosa2026}, and a weak-to-strong transition of the dissipation drives exactly this growth~\cite{ItoRamos2025}. Such a transient enhancement, together with the black holes and gravitational waves it produces, was computed in Ref.~\cite{Ballesteros2022} by evolving the covariance of the stochastic perturbations deterministically, and we use the same method. In our benchmark the CMB scales leave the horizon with $Q\equiv\Upsilon/3H\sim10^{-11}$ and behave as in cold inflation; the dissipation becomes strong only about twenty-four e-folds before the end. The question we ask is how the coupled inflaton--radiation system responds when its friction is driven in time, and we answer it by measurement. A small, localized change of the dissipation is inserted into the coupled solver and the power of a fixed mode is recomputed; repeating this at different insertion times gives a power-response kernel. This kernel is our main result.When we convolve this kernel with the periodic modulation of the friction, it reproduces the log-periodic comb in the full spectrum. The comb is therefore the linear response of the coupled system to the oscillating friction. Its integral, the static response, lies below the estimate built from published growth-function fits; our solver reproduces their slope, and the difference comes from where the thermal noise enters and how it is weighted. Within the tested window the modulus of the kernel follows a closed-form filter derived from a reduced freeze-out model.

The comb carries over into the induced gravitational waves. Log-periodic scalar features are known to survive there, with components at the original and the doubled frequency~\cite{Fumagalli2021} and a pattern that depends on the expansion history~\cite{Witkowski2022}. Log-periodic backgrounds have also been parametrized independently of their origin, for discrete scale invariance beyond Einstein gravity~\cite{CalcagniKuroyanagi2024}, and near-critical collapse can make the black-hole mass function itself log-periodic~\cite{Padilla2026,Padilla2026b}; our mechanism differs from both. Periodic structure can also amplify small-scale power resonantly and form black holes, through an oscillating sound speed~\cite{CaiTongWangYan2018} or a resonant instability during inflation~\cite{Zhou2020}; our comb is a modest, filtered modulation instead. A localized enhancement from a second mediator of the same sector, independent of the periodicity, gives a peaked feature that we use to illustrate black-hole formation and a millihertz gravitational-wave peak. Finally, the modulation reaches the bispectrum. For resonant features of the potential the oscillatory bispectrum is tied to the power spectrum beyond the squeezed limit, but no such relation is known for a modulated friction, so we compute the bispectrum at tree level.

The dissipation coefficient is a positive phenomenological ansatz with a periodic dependence on the field, and the background and the linear perturbations are computed without slow-roll or quasi-stationary approximations. The comb and its gravitational-wave imprint are linear-order results, the black-hole abundance and the millihertz peak are conditional Gaussian estimates, and the bispectrum is a tree-level result with the metric unperturbed. Section~\ref{sec:dissipation} sets up the periodic friction, Sec.~\ref{sec:background} the benchmark, and Sec.~\ref{sec:power} the spectrum and its response. Sections~\ref{sec:pbh}, \ref{sec:sigw} and \ref{sec:bispectrum} treat the black holes, the gravitational waves and the bispectrum, and Sec.~\ref{sec:conclusions} discusses the results and their limits. We use $M_{\rm Pl}=1$ and $g_*=106.75$.

\section{A periodic friction}
\label{sec:dissipation}

When the inflaton dissipates through a heavy mediator field that decays into light degrees of freedom, and the mediator is heavier than the temperature, the dissipation coefficient is cubic in the temperature of the bath~\cite{BGRR2013,BGRR2011diss},
\begin{equation}
\Upsilon(\phi,T)=C_\Upsilon\,\frac{g(\phi)^4\,T^3}{M_\chi(\phi)^2},
\label{eq:Upsilon}
\end{equation}
with $g$ the inflaton--mediator coupling and $M_\chi$ the mediator mass. Direct couplings to light fermions, as in the Warm Little Inflaton, give a coefficient linear in $T$ instead~\cite{WLI2016}, and minimal warm inflation has the cubic power through sphaleron heating~\cite{MinimalWarmInflation}. The derivations of the cubic law rest on specific interaction structures, and we do not derive it for the sector below: throughout, Eq.~\eqref{eq:Upsilon} is a positive phenomenological ansatz. If both $g$ and $M_\chi$ depend on $\phi/f$, keeping the first harmonic,
\begin{equation}
g=g_0\big[1+\alpha\cos(\phi/f)\big],\qquad
M_\chi^2=M_0^2\big[1+\beta\cos(\phi/f)\big],
\label{eq:couplings}
\end{equation}
the friction at fixed temperature, $A(\phi)\equiv\Upsilon/T^3$, is modulated to first order as
\begin{equation}
\delta\ln A=\delta_A\cos(\phi/f),\qquad \delta_A\equiv4\alpha-\beta .
\label{eq:bare}
\end{equation}
We carry $\delta_A=0.2$ as the single parameter of the modulation; it corresponds to $\alpha$ and $\beta$ of a few to twenty percent. The same sector underlies warm baryogenesis and dark-matter production~\cite{WarmBaryogenesis2012,RosaVentura2019}, so a mild modulation of its couplings adds little to an already motivated structure.

The perturbations do not respond to $A$ alone. The temperature adjusts to the friction through the radiation balance $4H\rho_r\simeq\Upsilon\dot\phi^2$, and the roll adjusts through $\dot\phi\simeq-V'/[3H(1+Q)]$, so the modulation of $Q$ is a dressed and possibly delayed version of $\delta_A$. A static change of $A$ on the attractor obeys $Q(1+Q)^6\propto A^4$, so that $d\ln Q/d\ln A=4/[1+6Q/(1+Q)]\simeq0.71$ at $Q\simeq3.3$; the integrated background gives $0.717$, within $0.3\%$ of this estimate. At the modulation frequency of the benchmark the response is larger, $\delta_Q=0.161$ or $\delta_Q/\delta_A=0.81$, and ahead in phase by $0.23$~rad, because the temperature follows the drive with a delay and only partly cancels it. This is why we measure the response of the spectrum instead of assuming it.

The modulation imprints an oscillation in $\ln k$ because the phase $\phi/f$ advances as modes leave the horizon. Its frequency per e-fold, and per unit $\ln k$ to leading slow-roll order, is
\begin{equation}
\omega=\frac1f\left|\frac{d\phi}{dN}\right|.
\label{eq:omega}
\end{equation}
We take $\omega$ positive. The field decreases along the roll, so the coupling phase $\theta=(\phi-\phi_c)/f$ decreases as successive modes cross, $d\theta/d\ln k\simeq-\omega$. In the benchmark $\omega\simeq1.9$ across the window of strong dissipation (between $1.7$ and $2.2$), about one cycle per $\Delta\ln k\simeq3.3$. That is fast enough to fit about three cycles in the window and slow enough to survive the filtering by freeze-out computed in Sec.~\ref{sec:power}.

A periodic mediator mass arises naturally from a mediator with a bare and an axion-induced mass. Take a Dirac fermion $\chi$ with a Yukawa coupling $y$ to a light bath scalar $\sigma$,
\begin{equation}
\mathcal{L}_\chi=i\,\bar\chi\gamma^\mu\partial_\mu\chi-\bar\chi\big[m_b-m_a\,e^{i\gamma_5\phi/f}\big]\chi-y\,\sigma\,\bar\chi\chi .
\label{eq:Lchi}
\end{equation}
The physical mass is the modulus of the complex mass, so its square is a pure first harmonic,
\begin{equation}
\begin{aligned}
M_\chi^2(\phi)&=m_b^2+m_a^2-2m_bm_a\cos(\phi/f)\\
&\equiv M_\infty^2\big[1-\varrho\cos(\phi/f)\big],
\end{aligned}
\label{eq:Mchifull}
\end{equation}
with $M_\infty^2=m_b^2+m_a^2$ and $\varrho=2m_bm_a/(m_b^2+m_a^2)$. A real mass term $m_b-m_a\cos(\phi/f)$ would get this wrong: its square drops the constant $m_a^2$ and adds a second harmonic. In Eq.~\eqref{eq:Lchi} the inflaton couples to the mediator only through this phase, so the effective coupling depends on $\phi/f$ as well, with a scalar part proportional to $\sin(\phi/f)$ and a pseudoscalar part proportional to $\cos(\phi/f)$ in the basis of Eq.~\eqref{eq:Lchi}. The modulation $\delta_A$ of the friction therefore combines a periodic mass and a periodic coupling and is not fixed by $\varrho$ alone. We have not derived the cubic law, or its coefficient, for a fermionic mediator, so the Lagrangian motivates the periodic dependence without realizing Eq.~\eqref{eq:Upsilon}, and $\delta_A$ remains a free parameter. Three further remarks about symmetry apply. The discrete periodicity of the couplings is what we use; it allows higher harmonics, which we neglect. A continuous shift symmetry, softly broken, is what would keep a potential flat, and our plateau breaks it explicitly. And a strict axion cannot sustain the strong regime through sphaleron heating~\cite{Zell2025}, so we do not identify the potential with an axion potential, and a large effective field range of the kind obtained from aligned sub-Planckian decay constants~\cite{KimNillesPeloso2005} is not needed.

The cubic law requires the mediator to be heavier than the temperature, and this needs care exactly where the friction surges. Writing $\Upsilon=C_{\rm eff}\,T^3/M_\chi^2$ as $A(\phi)T^3$ defines the mediator mass that the friction implies, $M_\chi^2=C_{\rm eff}/A(\phi)$, and along the trajectory
\begin{equation}
\Big(\frac{M_\chi}{T}\Big)^2=\frac{C_{\rm eff}}{3Q}\,\frac{T}{H}.
\label{eq:MoverT}
\end{equation}
For $C_{\rm eff}=10^3$ the ratio never falls below $26.7$ in the benchmark, with its minimum at the peak of the localized enhancement ($Q=9.05$, $T/H=19.3$; Fig.~\ref{fig:background}). Keeping $M_\chi/T\geq10$ throughout requires $C_{\rm eff}\gtrsim1.4\times10^2$. This is a condition on a free coefficient, not a check of a derived one: $C_{\rm eff}$ grows with the couplings and the multiplicities of the sector, and strong dissipation at low temperature is known to require large multiplicities~\cite{BGRR2011diss,BGRR2013}. With one common coefficient for the whole of $A(\phi)$ this is a proxy; a realization with separate mediators would have one such condition for each. It constrains masses, not thermal equilibrium, which we examine in Sec.~\ref{sec:background}.

The same sector also generates a periodic potential. At one loop a Dirac mediator with the mass of Eq.~\eqref{eq:Mchifull} contributes to $V$ a term $\varrho M_\infty^4\cos(\phi/f)/8\pi^2$ at a renormalization scale $M_\infty$~\cite{ColemanWeinberg1973}. With $M_\infty^2=C_{\rm eff}/A$ in the window, the slope of this term is about three times the slope of the plateau there for $C_{\rm eff}=1.4\times10^2$, and about $170$ times for $C_{\rm eff}=10^3$, so a fermionic mediator on its own is excluded. In the supersymmetric two-stage mechanism in which the low-temperature cubic law was obtained~\cite{BereraMossRamos2009,BGRR2011diss}, the scalar partners cancel this loop up to the splitting that breaks supersymmetry, and the remainder can be estimated. Let the mediators be $N_X$ pairs of chiral multiplets with the superpotential $\mu(\Phi)X\bar X$ and $\mu=m_b-m_a e^{i\Phi/f}$, so that the fermion has the mass of Eq.~\eqref{eq:Mchifull}. If the vacuum energy of the plateau is the $F$-term of the inflaton multiplet, $|F_\Phi|^2=V$, the cross term of $|\partial W/\partial\Phi|^2$ mixes the scalars as $Bx\bar x+{\rm c.c.}$ with $|B|=|\mu'||F_\Phi|=m_aV^{1/2}/f\equiv\Delta$. Their masses are split to $|\mu|^2\pm\Delta$ while the fermion stays at $|\mu|^2$, so that per pair ${\rm STr}\,\mathcal M^2=0$ and ${\rm STr}\,\mathcal M^4=4\Delta^2$. Expanding the Coleman--Weinberg potential in $x=\Delta/|\mu|^2$, with $(1+x)^2\ln(1+x)+(1-x)^2\ln(1-x)=3x^2-x^4/6+\dots$, the constant terms cancel and
\begin{equation}
V_1=\frac{N_X\Delta^2}{16\pi^2}\ln\frac{|\mu|^2}{\Lambda^2}-\frac{N_X\Delta^4}{192\pi^2|\mu|^4}+\dots
\label{eq:Vsusy}
\end{equation}
The renormalization scale $\Lambda$ multiplies $\Delta^2$, which does not depend on $\phi/f$, so the periodic part is independent of it. Here $\Delta/|\mu|^2\leq0.06$. The first harmonic of $\ln(1-\varrho\cos\theta)$ is $-\varrho(1+\varrho^2/4+\dots)$, and the first harmonic of the full one-loop expression, extracted numerically, agrees with the leading term to $1\%$.

The periodic part of Eq.~\eqref{eq:Vsusy},
\[
V_1^{\rm per}=-\frac{N_X\varrho\,m_a^2V}{16\pi^2f^2}\cos(\phi/f),
\]
has a slope $b=3.5\times10^{-3}N_X$ times that of the plateau for $C_{\rm eff}=1.4\times10^2$ and $b=2.5\times10^{-2}N_X$ for $C_{\rm eff}=10^3$, taking $\varrho=0.2$, the size of $\delta_A$, here and above. These numbers assume the largest splitting, from the $F$-term of the inflaton multiplet. If the vacuum energy resides in another multiplet, as in many supergravity models of inflation, Hubble-induced masses of order $H^2$ split the scalars instead. A common shift $\delta m^2$ of the scalar masses gives ${\rm STr}\,\mathcal M^4=8|\mu|^2\delta m^2$ per pair, and for $\delta m^2=H^2$, with the logarithm run from the Planck scale, the remainder is about twenty times smaller. The bath breaks supersymmetry as well, but with $M_\chi/T\geq10$ the mediators are Boltzmann suppressed, and the light fields of the bath couple to the mediator loop through self-energies that supersymmetry protects at zero momentum; we have not computed these thermal terms.

In cold slow roll a slope modulated by a fraction $b$ modulates the power by a fraction between $2b$, the static value, and $3b(2\pi/\omega)^{1/2}\simeq5b$, the resonant estimate at $\omega\simeq1.9$~\cite{FlaugerOscillations2010,FlaugerPajer2011}. If a coefficient of this size holds with friction, which we have not computed, the potential adds to the comb an oscillation of the cold-inflation kind of $2$--$6\%$ of $\hat A_1\simeq0.34$ per pair for $C_{\rm eff}=1.4\times10^2$, and $15$--$40\%$ per pair for $C_{\rm eff}=10^3$. The dissipative comb therefore dominates if $C_{\rm eff}$ stays near its lower bound and the mediators are few. Since $C_{\rm eff}$ grows with the multiplicities of both the mediators and their decay products, while the potential grows with $N_X$ alone, the large multiplicity that strong dissipation requires~\cite{BGRR2011diss,BGRR2013} should reside in the decay products. The multiplicities and perturbative couplings that a complete realization must supply are not derived here.

\section{Background and benchmark}
\label{sec:background}

The CMB constrains the potential before anything else. Natural inflation, $V=\Lambda^4[1+\cos(\phi/f_{\rm N})]$ with $f_{\rm N}=5$ and $N_*=55$, gives $n_s\simeq0.950$ and $r\simeq0.039$, which the data exclude. We therefore place the periodic friction on a plateau with a quadratic minimum,
\begin{equation}
V(\phi)=V_0\,\frac{\phi^2}{\phi^2+\mu^2},
\label{eq:Vnat}
\end{equation}
and let the cubic coefficient carry the periodic factor, a localized enhancement and a broad window of strong dissipation later in the roll,
\begin{equation}
\begin{aligned}\Upsilon=T^3\Big\{&A_{\rm bg}+\big[A_b\,e^{-(\phi-\phi_b)^2/2w^2}+A_w\,S_w(\phi)\big]\\&\times\big[1+\delta_A\cos\!\big((\phi-\phi_c)/f\big)\big]\Big\},\end{aligned}
\label{eq:Upsmod}
\end{equation}
\begin{equation}
S_w(\phi)=\tfrac12\Big[\tanh\frac{\phi-\phi_c+h}{\Delta_t}-\tanh\frac{\phi-\phi_c-h}{\Delta_t}\Big].
\label{eq:Sw}
\end{equation}
The Gaussian term stands for a single mediator that becomes light near an enhanced-symmetry point $\phi_b$. Near its centre it matches a quadratic mass crossing, $M_\chi^2=m_0^2[1+(\phi-\phi_b)^2/\Delta^2]$ with $w=\Delta/\sqrt2$, but it falls off faster; with everything else fixed, the full Lorentzian lengthens inflation from $97.7$ to $102.0$ e-folds. Its tails, which fall only as the inverse square of the distance, would also raise $A$ at the pivot by a factor of $2.7$ and $Q_*$ by about $50$, still below $10^{-9}$, and they would move the feature in $N$, which would then need retuning; we keep the Gaussian as a phenomenological profile that confines the enhancement to the crossing. The window would correspond to a second mediator that stays light over a field range $2h$. Both envelopes are phenomenological shapes. The resonance at $\phi_b$ is one crossing and not a tooth of the comb: the periodicity makes the couplings oscillate, while the enhancement comes from a mediator becoming light. The background coefficient $A_{\rm bg}$ carries no periodic factor; it sets only the negligible friction at CMB scales, where a modulation would be unobservable. The inputs are listed in Table~\ref{tab:bench}. We calibrate to $A_s=2.1\times10^{-9}$ at the pivot by rescaling $V_0\to\varsigma V_0$ and every coefficient $A\to\varsigma^{-1/4}A$ together. This leaves $\phi(N)$, $Q$ and $\rho_r/V$ unchanged, but $T/H$ scales as $\varsigma^{-1/4}$ and with it the noise and $M_\chi/T$, so every spectrum below is computed on the calibrated background.

The background obeys
\begin{align}
3H^2&=V+\rho_r+\tfrac12\dot\phi^2,\label{eq:bgH}\\
\ddot\phi+(3H+\Upsilon)\dot\phi+V'&=0,\label{eq:bgphi}\\
\dot\rho_r+4H\rho_r&=\Upsilon\dot\phi^2,\label{eq:bgrho}
\end{align}
closed by $T=(30\rho_r/\pi^2g_*)^{1/4}$. We integrate these equations in full, keeping $\ddot\phi$ and $\dot\rho_r$, from $\phi_0=5$ to $\epsilon_H=1$. The field starts on the attractor $\dot\phi\simeq-V'/[3H(1+Q)]$, $\rho_r\simeq\Upsilon\dot\phi^2/4H$, with $Q$ solved self-consistently. Inflation lasts $97.73$ e-folds, and the pivot $k_*=0.05\Mpc^{-1}$ leaves the horizon $42.7$ e-folds after the start and $55$ before the end.

\begin{table}[t]
\caption{The benchmark after calibration, in reduced Planck units. The upper block lists the inputs to the precision used, the lower block derived quantities.}
\label{tab:bench}
\footnotesize\setlength{\tabcolsep}{4pt}
\begin{tabular}{ll}
\hline\noalign{\smallskip}
$V_0,\ \mu$ & $1.7422\times10^{-10},\ 1$\\
$A_{\rm bg}$ & $8704.1$\\
$A_b,\ A_w$ & $8.3995\times10^{7},\ 1.30562\times10^{7}$\\
$\phi_b,\ w$ & $3.09,\ 0.01$\\
$\phi_c,\ h,\ \Delta_t$ & $2.965,\ 0.077,\ 0.004$\\
$f,\ \delta_A$ & $8.3\times10^{-3},\ 0.2$\\
\noalign{\smallskip}\hline\noalign{\smallskip}
pivot $\phi_*,\ d\phi/dN$ & $4.1615,\ -0.0261$\\
pivot $\epsilon_H,\ Q_*$ & $3.4\times10^{-4},\ 1.1\times10^{-11}$\\
pivot $T_*/H_*$ & $0.041$\\
$n_s,\ r$ & $0.964,\ 5.3\times10^{-3}$\\
feature $Q_{\rm pk},\ T/H$ & $9.05,\ 19.3$\\
feature $\max|\eta_H|$ & $2.9$\\
window e-folds with $Q>1$ & $9.4$\\
window $\omega$ & $1.9$ ($1.7$--$2.2$)\\
end $\phi,\ Q,\ T/H$ & $0.543,\ 3.3\times10^{-3},\ 38$\\
\noalign{\smallskip}\hline
\end{tabular}
\end{table}

At the pivot the dissipation is negligible, $Q_*\simeq1.1\times10^{-11}$ with $T_*/H_*\simeq0.04$, and the CMB scales are those of cold inflation on the plateau. Mode integration gives $n_s=0.964$ and $r=5.3\times10^{-3}$, and switching on the thermal noise changes the pivot power by a fraction $5\times10^{-15}$. Placing the pivot in the weak regime is a choice. Cubic dissipation makes the growth function of Sec.~\ref{sec:power} rise steeply with $Q$, so a strongly dissipative pivot on a trajectory along which $Q$ grows would tend to give a blue tilt. The benchmark agrees with the Planck value $n_s=0.9649\pm0.0042$~\cite{Planck2018X} at $0.2\sigma$ and with the BICEP/Keck bound $r<0.036$~\cite{BK18}. The P-ACT-LB combination, which adds ACT DR6, CMB lensing and DESI baryon acoustic oscillations to Planck, prefers $n_s=0.974\pm0.003$~\cite{ACTDR6}, about three standard deviations higher. The reason is simple: e-folds spent in strong dissipation cover little field range, so the feature and the window, which together hold $Q>1$ for about fifteen e-folds, push the pivot down the plateau. The tilt, however, is set by the plateau near $\phi_*$, far above the field values of the dissipative sector, and it can be changed there without touching that sector. Adding $V_0c\,(\phi-\phi_t)^3$ for $\phi>\phi_t=3.6$, with $c=1.35\times10^{-3}$, keeps $V$ and its first two derivatives continuous and leaves everything below $\phi_t$ alone. Recalibrated, the deformed plateau gives $n_s=0.974$, $r=5.9\times10^{-3}$ and $dn_s/d\ln k=-1.3\times10^{-3}$ ($-8.4\times10^{-4}$ for the benchmark). The trajectory of $Q$ is unchanged; $V_0$ rises by $10.7\%$, and $T/H$ in the feature and the window falls by a factor $0.975$. Recomputed with the coupled solver, the comb fundamental of Sec.~\ref{sec:power} is $0.3421$ against $0.3424$ on the benchmark, and the window power rises by $7\%$. The feature is more sensitive: at fixed $A_b$ its peak power rises by $7\%$ and the illustrative dark-matter fraction of Sec.~\ref{sec:pbh} from $0.53$ to $3.8$, while lowering $A_b$ by $1.5\%$ restores $0.54$ at the same black-hole mass. The comb is therefore the same at either tilt, and the black holes need only a percent-level retuning. Neither model has been fitted to a full likelihood, and we keep the undeformed plateau as the benchmark for the rest of the paper.

\begin{figure*}[t]
\centering
\includegraphics[width=\textwidth]{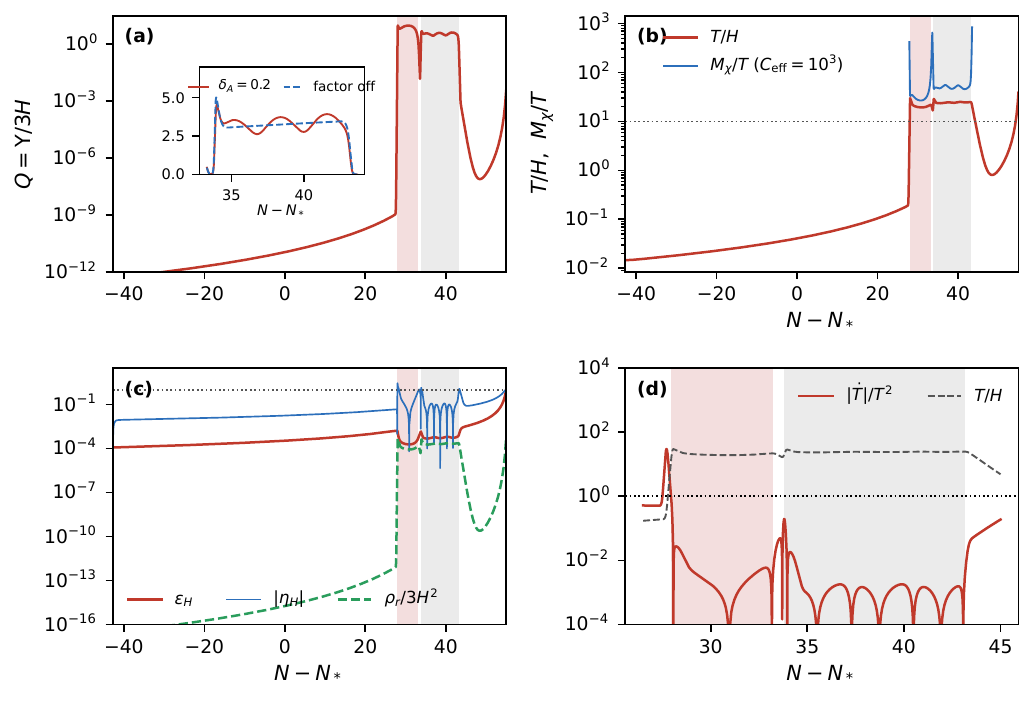}
\caption{The benchmark background against e-folds from the CMB pivot, with the feature (pink) and the window (grey) shaded where $Q>1$. (a) The dissipation ratio $Q=\Upsilon/3H$, with the periodic factor of the window on (red) and off (blue dashed); the inset shows the window on a linear scale. (b) The bath ratio $T/H$ and the mediator-mass proxy $M_\chi/T$ of Eq.~\eqref{eq:MoverT} for $C_{\rm eff}=10^3$ (dotted line: $M_\chi/T=10$). (c) $\epsilon_H$, $|\eta_H|$ and $\rho_r/3H^2$; inflation continues through the feature although $|\eta_H|$ exceeds one at its onset. (d) The heating rate of the bath in units of its temperature, $|\dot T|/T^2$, through the onset of the feature and the window, with $T/H$ (dashed) for reference. A weakly coupled bath cannot keep up above the dotted line at one.}
\label{fig:background}
\end{figure*}

The dissipation becomes strong late. At the onset of the feature, about twenty-four e-folds before the end, $Q$ rises from below $10^{-2}$ to $9.05$ within $0.16$ e-fold, and the bath is heated from $T/H\simeq0.2$ to $19$ (Fig.~\ref{fig:background}). Modes leaving the horizon about two e-folds after $Q$ peaks form the curvature peak, $\Pz=6.83\times10^{-3}$ at $\kpk=1.18\times10^{13}\Mpc^{-1}$, $\ln(\kpk/k_*)=33.1$. Here and below $\kpk$ is the maximum of the spectrum, located by a parabola through the three largest samples on a grid of $0.2$ e-fold; halving the grid moves it by $0.2\%$. The peak is placed, not predicted: its position is fixed by $\phi_b$ and its width by $w$. The window opens $0.6$ e-fold after the strong phase of the feature ends, holds $Q>1$ for $9.4$ e-folds with $Q\simeq2.6$--$3.9$ in its interior, and closes $11.8$ e-folds before the end of inflation. The modulation frequency drifts as the field accelerates, much like the drifting oscillations of monodromy models~\cite{FlaugerDrifting2017}. Across the feature, where the field nearly stalls, $\omega$ runs from $0.7$ to $6.7$ and the coupling completes less than a cycle, so the periodic factor only reshapes the peak. Across the window $\omega$ stays near $1.9$, and the coupling completes $2.9$ cycles over $\Delta\ln k\simeq9.4$. The window is where the comb lives. Its curvature power sits on a plateau $\Pz\simeq(0.5$--$1.1)\times10^{-4}$, far below any collapse threshold, so only the feature can form black holes. To isolate the modulation we also integrate a reference trajectory with the periodic factor of the window set to one and everything else unchanged, recalibrated to the same $A_s$ (which raises $V_0$ by $0.12\%$). Compared at the same field value, $\ln(Q/Q_{\rm ref})\simeq\delta_Q\cos(\theta-0.23)$ with the $\delta_Q=0.161$ of Sec.~\ref{sec:dissipation}; since $\theta$ decreases in time, the maxima of $Q$ come before those of $A$, which is the lead quoted there.

The surge neither traps the field nor ends inflation. Through the feature and the window $\epsilon_H$ stays below $1.8\times10^{-3}$ and $\rho_r/3H^2$ below $5\times10^{-4}$. The second slow-roll parameter, $\eta_H=\epsilon_H-\tfrac12d\ln\epsilon_H/dN$, does not stay small: it reaches $|\eta_H|=2.9$ at the onset of the feature and $1.4$ at the edges of the window, while it stays below $0.12$ inside the window. Inflation persists through this transient violation of slow roll, as it does through the oscillations of $\eta$ in the oscillating-mass warm inflaton~\cite{FerrazRosa2023}, and since we solve the full background and linear equations no slow-roll expansion is needed. Inflation ends at $\phi=0.543$ with $Q=3.3\times10^{-3}$, $\rho_r/3H^2=5\times10^{-4}$ and $T/H=38$, so the bath is warm but far from dominant. If the inflaton then oscillates about its quadratic minimum and decays, instant thermalization would correspond to $N_*=55.8$; keeping $N_*=55$ requires a matter-like phase of $3.3$ e-folds, which ends at $T_{\rm reh}\simeq2\times10^{14}$~GeV. A different reheating history moves $N_*$, and with it the scale of the feature and the calibration of $V_0$; for the black holes this is absorbed into the choice of $A_b$. Every mode used below re-enters at $T\leq3.4\times10^{11}$~GeV, so the radiation-era conversions of Secs.~\ref{sec:pbh} and \ref{sec:sigw} hold for any reheating that completes above that temperature.

A temperature can describe the bath only if the bath relaxes faster than it is heated, and neither a large $T/H$ nor a large $M_\chi/T$ guarantees this~\cite{BGRR2018Boltzmann}. A useful measure is the heating rate in units of the temperature, $|\dot T|/T^2=|d\ln T/dN|\,H/T$, since a weakly coupled plasma relaxes at a rate below $T$. Through the onset of the feature this ratio reaches $30$, when $T/H\simeq0.36$ and the bath is still forming, and it stays above one for $0.45$ e-fold while $T/H$ climbs from $0.2$ to $14$ (Fig.~\ref{fig:background}). The thermal description of the onset is therefore not under control, and fixing the amplitude of the feature would need a kinetic treatment of the bath. In the window the ratio stays below $0.2$ at the edges and below $2\times10^{-3}$ in the interior. In units of the expansion rate the bath is heated at $|d\ln T/dN|\leq0.05$ in the interior and up to $4$ at the edges, so a thermal window needs a relaxation rate $\Gamma_{\rm rel}\gg H$ inside and $\Gamma_{\rm rel}\gtrsim4H$ at the edges, that is $\Gamma_{\rm rel}/T\gg0.04$ and $\gtrsim0.2$. We do not compute $\Gamma_{\rm rel}$, which depends on couplings that the phenomenological coefficient leaves open, so a thermal window is an assumption of this paper; Sec.~\ref{sec:coupled} shows how much the comb changes if the bath relaxes at a finite rate. The window also inherits its bath from the feature, with $T/H\geq16.7$ in between. Removing the feature does not help: on a window-only background the window opens abruptly and $|\dot T|/T^2$ reaches about $33$. The comb fundamental is nevertheless $0.344$ without the feature and $0.342$ with it, so under the noise closure of Sec.~\ref{sec:power} the comb does not depend on how the window is prepared, even though neither preparation is shown to be thermal. The boundary between quantum- and thermal-dominated fluctuations is not the line $T/H=1$~\cite{RamosdaSilva2013}, and throughout the window $T/H\simeq23$--$25$.

\section{The curvature spectrum and its response}
\label{sec:power}

\subsection{Coupled perturbations}
\label{sec:coupled}

How fast the background varies decides how the spectrum must be computed. Across the window the friction changes slowly, with $|d\ln(1+Q)/dN|\simeq0.14$ (median), against the geometric rate of about two that sets the width of freeze-out. At the onset of the feature the same rate reaches $27$, and no quasi-stationary or overdamped reduction can be trusted there. We therefore compute the spectrum everywhere from the coupled linear perturbations of the inflaton, the radiation and the metric. In Newtonian gauge, $ds^2=-(1+2\Psi)dt^2+a^2(1-2\Psi)d\bm{x}^2$, with $u=\dot\phi$, radiation momentum $\delta T^0{}_i=\partial_iq_r$ and $\delta\Upsilon=\Upsilon_{,\phi}\delta\phi+(3\Upsilon/4\rho_r)\delta\rho_r$, they read~\cite{BGRR2014fluid,Ballesteros2022}
\begin{align}
&\delta\ddot\phi+(3H+\Upsilon)\delta\dot\phi+\Big(\tfrac{k^2}{a^2}+V''+u\Upsilon_{,\phi}\Big)\delta\phi+\tfrac{3\Upsilon u}{4\rho_r}\delta\rho_r\nonumber\\
&\qquad=4u\dot\Psi-(2V'+\Upsilon u)\Psi+\xi,\label{eq:cdphi}\\
&\delta\dot\rho_r+\Big(4H-\tfrac{3\Upsilon u^2}{4\rho_r}\Big)\delta\rho_r-\tfrac{k^2}{a^2}q_r\nonumber\\
&\qquad=4\rho_r\dot\Psi+u^2\Upsilon_{,\phi}\delta\phi+2\Upsilon u\,\delta\dot\phi-\Upsilon u^2\Psi-u\,\xi,\label{eq:cdrho}\\
&\dot q_r+3Hq_r+\tfrac13\delta\rho_r+\tfrac43\rho_r\Psi=-\Upsilon u\,\delta\phi,\label{eq:cdq}\\
&\dot\Psi+H\Psi=\tfrac12(u\,\delta\phi-q_r).\label{eq:cdPsi}
\end{align}
The radiation is an ideal fluid without anisotropic stress, and the noise $\xi$ is Markovian. Shear viscosity acts as an extra friction on the radiation fluctuations and can suppress the growing mode~\cite{ShearViscous2011}, so the ideal fluid gives the largest coupling-driven growth; we have not scanned that freedom. The noise enters the two energy equations with opposite signs, so the exchange conserves energy, and the energy constraint, with Eq.~\eqref{eq:cdPsi} substituted, decays as $a^{-3}$ identically, the stochastic source included. For each $k$ the vector $X=k^{3/2}(\delta\phi,\delta\dot\phi/H,\delta\rho_r/H^2,q_r/H,\Psi)$ obeys a linear system $dX=\mathsf{A}X\,dN+b\,dW_N$, and we evolve its covariance exactly, without Monte Carlo sampling,
\begin{equation}
\begin{aligned}
\Sigma'&=\mathsf{A}\Sigma+\Sigma\mathsf{A}^{\rm T}+bb^{\rm T},\\
b&=H\sqrt{6Q\,(T/H)\,z^3S(\tilde z)}\,(0,1,-v,0,0)^{\rm T},
\end{aligned}
\label{eq:cov}
\end{equation}
with $z=k/aH$ and $v=d\phi/dN$. The spectrum $\Pz$ is that of the comoving curvature perturbation of the total fluid, $\mathcal R=\Psi+H(u\,\delta\phi-q_r)/(u^2+\tfrac43\rho_r)$, not of the inflaton alone. The noise is weighted by the occupation of the bath,
\begin{equation}
S(\tilde z)=\frac{\tilde z}{e^{\tilde z}-1},\qquad \tilde z=\frac{k/a}{T}=\frac{zH}{T},
\label{eq:occ}
\end{equation}
a weighting obtained by matching the noise correlator to equilibrium quantum statistics deep inside the horizon~\cite{LaineProcacciRogelj2025}. The symmetric correlator of a thermal mode is $\omega\coth(\omega/2T)=\omega+2\omega n_B(\omega)$. We let only the thermal part, $2\Upsilon TS(\tilde z)$, drive the stochastic term and keep the zero-point part in the initial vacuum covariance; white noise of strength $2\Upsilon T$ at all momenta overcounts the source for $k/a>T$. When $\Upsilon\gtrsim H$ the friction damps that initial vacuum before horizon crossing, so the prescription drops the vacuum fluctuation that a zero-point noise would sustain. Relative to the thermal term of the bracket $F$ of Sec.~\ref{sec:static}, about $130$ in the window, this is below one percent there, but it is not small at the onset of the feature, where $T/H$ is of order one or less. In the matching we replace the mode energy by $k/a$, neglecting the dissipative width, which is modest in the window ($\Upsilon/T\simeq0.4$) and not small at the feature ($\simeq1.4$).

This closure is not derived for our sector, so everything that depends on it was recomputed with four alternatives: white noise, a Gaussian cutoff $e^{-\tilde z^2/2}$, $S(\tilde z/2)$ and $S(2\tilde z)$. They change the power at the black-hole peak by factors $1.62$, $1.07$, $1.28$ and $0.59$, and the window plateau by $1.28$, $1.13$, $1.13$ and $0.77$. Among the approximations we have tested, the closure and, for slow relaxation, a finite memory of the noise (below) dominate the uncertainty of the linear amplitude, at the level of tens of percent; viscosity and the non-thermal onset have not been estimated. Modes start in the vacuum at $z=100$, where $k/a\simeq4T$ in the window and the radiation is not yet hydrodynamic; the occupation weighting suppresses the noise there, and an earlier start changes little, as shown below. They are followed seven e-folds past horizon crossing, with an implicit Radau integrator. A tighter tolerance, a start at $z=180$, nine e-folds past crossing, or a halved maximum step change the power by at most $4\times10^{-7}$, $3\times10^{-5}$, $7\times10^{-7}$ and $4\times10^{-7}$, at the feature and in the window alike, and the constraint holds to $3\times10^{-4}$ throughout the evolution. These tests bound the numerical error of the solver for the equations as stated. They do not test whether those equations describe the bath, that is, whether the radiation is an ideal fluid in thermal equilibrium and whether its noise is Markovian; those are physical approximations, which we probe separately with the closure scan above and the memory scan below, and otherwise state as assumptions.

The noise is Markovian only if the bath relaxes much faster than the modes evolve. To see how much this matters for the comb, we replaced the Markovian noise by an Ornstein--Uhlenbeck noise of correlation time $\tau_c$, normalized so that its correlation integrates to the Markovian strength, and evolved the enlarged covariance exactly; as $\tau_c\to0$ it reproduces the Markovian power to $10^{-4}$. We recomputed the window modes of the benchmark, with and without modulation, for $H\tau_c=0.03$, $0.1$ and $0.3$, that is for $(3H+\Upsilon)\tau_c=0.4$, $1.3$ and $4.0$, or a relaxation rate $\Gamma_{\rm rel}=1/\tau_c$ of $33H$, $10H$ and $3.3H$ ($\Gamma_{\rm rel}/T=1.3$, $0.39$ and $0.13$). The window power falls by $1\%$, $7\%$ and $27\%$, while the comb fundamental changes by $-0.2\%$, $-1.2\%$ and $-3.5\%$ and its phase by at most $0.02$~rad. The comb is a ratio, so a finite memory, like the closure, moves the power much more than the comb.

\subsection{Static response}
\label{sec:static}

On a slowly varying background the warm spectrum is usually written $\Pz=(H^2/2\pi\dot\phi)^2F\,G(Q)$~\cite{BGRR2014fluid,BenettiRamos2017,Montefalcone2024}. The bracket $F=1+2n+2\sqrt3\pi Q\,(T/H)/\sqrt{3+4\pi Q}$ is the analytic result for an inflaton driven by thermal noise~\cite{GrahamMoss2009,BGRR2014fluid,Visinelli2015,ShearViscous2011,FlucDiss2018}, with $n$ the inflaton occupation, and the growth function $G(Q)$, fitted to numerical solutions, adds the growing mode produced by the coupling to radiation when $\Upsilon$ depends on $T$. The factors multiply, so their logarithmic slopes add,
\begin{equation}
p\equiv\frac{d\ln\Pz}{d\ln Q}=\frac{2Q}{1+Q}+\frac{d\ln F}{d\ln Q}+\frac{d\ln G}{d\ln Q}.
\label{eq:Q52}
\end{equation}
We take the derivative at fixed $V'$ and $H$, with $T/H\propto Q^{1/4}(1+Q)^{-1/2}$ from the radiation balance and $n=0$, as in the vacuum initial state of the solver; a thermal $n$ lowers $p$ by $0.1$--$0.15$. The first two terms, the kinematic prefactor, tend to $5/2$ at fixed $T$ and to $9/4$ once the radiation balance is included. The growth function adds a steep slope of its own. For cubic dissipation we use the two fits with documented provenance, the power-sum form of Benetti and Ramos~\cite{BenettiRamos2017} and the logarithmic fit of the WarmSPy analysis~\cite{Montefalcone2024},
\begin{equation}
\begin{aligned}
G_{\rm BR}(Q)&=1+4.981\,Q^{1.946}+0.127\,Q^{4.330},\\
G_{\rm W}(Q)&=10^{\,2.86x-0.18x^2+0.55x^3-0.0714x^4},\\
x&=\log_{10}(1+Q),
\end{aligned}
\label{eq:GQ}
\end{equation}
both used inside the range in which they were obtained. Their slopes are $2.5$--$2.7$ in the window and $3.5$--$3.9$ at the feature, so the full exponent is $p\simeq4.44$--$4.62$ in the window ($Q\simeq3.4$) and $5.6$--$6.0$ at the feature ($Q\simeq9$).

Equation~\eqref{eq:Q52} is an adiabatic estimate, and the coupled solver can test it. Changing $A$ at fixed field and recomputing the power gives the static response $S_A\equiv d\ln\Pz/d\ln A=2.71$ at $Q=3.28$ and $2.88$ at $Q=4.74$. Dividing by the measured $d\ln Q/d\ln A=0.717$ and $0.672$ gives $p=3.78$ and $4.29$, whereas Eq.~\eqref{eq:Q52} gives $4.40$--$4.56$ and $4.83$--$5.16$ at the same points, $12$--$21\%$ higher.

The difference lies in the noise, not in the solution. To see this we computed $G(Q)$ with our solver on de Sitter backgrounds with constant $Q$, $T/H$ and $\dot\phi$ on the slow-roll attractor ($d\phi/dN=10^{-3}$, $\rho_r/H^2=\tfrac34Q(d\phi/dN)^2$, $A_{,\phi}=0$), for $\Upsilon\propto T^c$ with white noise, $n=0$ and the vacuum at $z=100$, reading the power off seven e-folds after crossing; a start at $z=200$ or a read-off after ten e-folds changes $G$ by less than $0.6\%$. We define $G$ as the computed power for $c=3$ divided by $(H^2/2\pi\dot\phi)^2F$. For $c=0$ the same solver reproduces $F$ to within $2\%$ for $Q\geq1$ and $8\%$ below.

\begin{table}[t]
\caption{The growth function at $T/H=20$ computed on constant backgrounds, with the noise in the inflaton equation only and in the energy-conserving placement of Eq.~\eqref{eq:cdrho}, against the fits of Eq.~\eqref{eq:GQ}.}
\label{tab:G}
\footnotesize\setlength{\tabcolsep}{4pt}
\begin{tabular}{lcccc}
\hline\noalign{\smallskip}
$Q$ & inflaton only & energy-conserving & $G_{\rm BR}$ & $G_{\rm W}$\\
\noalign{\smallskip}\hline\noalign{\smallskip}
$0.1$ & $0.76$ & $17.4$ & $1.06$ & $1.31$\\
$1$ & $8.50$ & $28.5$ & $6.11$ & $7.23$\\
$3.4$ & $108$ & $171$ & $80.3$ & $79.4$\\
$9$ & $2.10\times10^3$ & $2.50\times10^3$ & $2.08\times10^3$ & $1.44\times10^3$\\
$20$ & $5.44\times10^4$ & $5.88\times10^4$ & $5.63\times10^4$ & $3.31\times10^4$\\
$100$ & $4.61\times10^8$ & $4.68\times10^8$ & $5.81\times10^7$ & $1.93\times10^8$\\
\noalign{\smallskip}\hline
\end{tabular}
\end{table}

With the noise in the inflaton equation only, as in the numerical work behind the fits~\cite{Montefalcone2024}, the computed $G$ lies within $40\%$ of $G_{\rm BR}$ for $1\leq Q\leq20$, a range over which the two fits differ from each other by up to $70\%$ (Table~\ref{tab:G}). Its slope is $2.56$ at $Q=3.4$ and $3.58$ at $Q=9$, inside the ranges of the fits, and for $Q\geq1$ it changes by less than $3\%$ between $T/H=10$ and $40$. Above $Q=20$ it rises faster than either fit, to $2.4$ times $G_{\rm W}$ at $Q=100$. Below $Q\simeq0.3$ it falls below one, because the feedback through $\delta\Upsilon$ first acts as an extra friction on $\delta\dot\phi$; a thermal inflaton occupation, which we do not include, would dominate $F$ there. Along the radiation balance, from pairs of constant backgrounds at $Q(1\pm0.02)$, it gives $p=4.46$ and $4.88$ at the two points, inside the ranges given by the fits.

Adding the noise to the radiation energy equation with the opposite sign, as in Eq.~\eqref{eq:cdrho}, lowers $p$ to $4.03$ and $4.56$. The Einstein equations do not decide whether the noise enters the energy or the momentum flux of the radiation, and Ref.~\cite{BGRR2014fluid} found that the first choice enhances the dissipative power. The radiation fluctuation that the noise creates decays outside the horizon at the rate $4-c$ and drives the inflaton through $\delta\Upsilon$. For $Q\ll1$ this multiplies the thermal power by $[4/(4-c)]^2=16$, close to the computed $17$--$18$ for $0.03\leq Q\leq0.3$. The enhancement falls with $Q$, to a factor $3.4$ at $Q=1$, $1.6$ at $Q=3.4$ and $1.2$ at $Q=9$, so it also flattens the growth. Weighting the noise by the occupation, Eq.~\eqref{eq:occ}, lowers $p$ further to $3.89$ and $4.38$, $3.0\%$ and $2.1\%$ above the static response; the remaining difference includes the slow evolution of the background, which constant backgrounds omit. The solver therefore reproduces the slope of the fits, and the fits overestimate the exponent here because they describe another placement and spectrum of the noise.

\begin{figure*}[t]
\centering
\includegraphics[width=\textwidth]{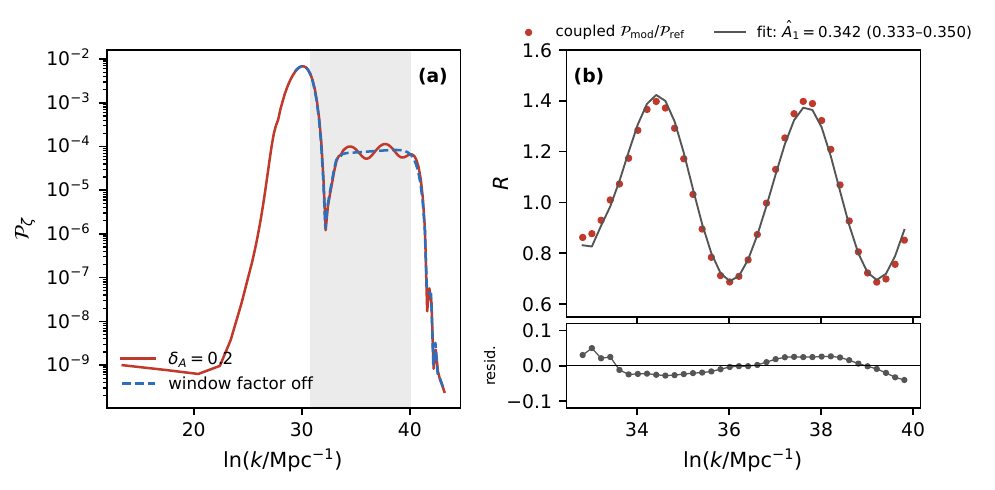}
\caption{The curvature spectrum and its comb. (a) $\Pz$ of the benchmark from the coupled perturbations, with the periodic factor of the window on (red) and off (blue dashed). The feature peaks at $\kpk=1.18\times10^{13}\Mpc^{-1}$ with $\Pz=6.83\times10^{-3}$; the grey band marks the modes that cross during the window. (b) The ratio $R=\mathcal P_{\rm mod}/\mathcal P_{\rm ref}$ on the fitted modes with the fit of Eq.~\eqref{eq:fitmodel}, $\hat A_1=0.342$, and below it the residual relative to the mean.}
\label{fig:spectrum}
\end{figure*}

From here on we use the measured static response $S_A$ instead of $p$. Because the spectrum responds to the friction with this exponent, a static modulation of $A$ enters the spectrum as
\begin{equation}
\Pz(k)=\mathcal{A}(k)\big[1+\delta_A\cos\theta(k)\big]^{S_A},
\label{eq:Pzeta}
\end{equation}
with $\mathcal A(k)$ the smooth envelope and $\theta(k)$ the coupling phase at freeze-out, for a modulation slow enough to be static. The bracket generates a zero mode and harmonics. About the mean, the fundamental is
\begin{equation}
a_1=s\,\delta\Big[1-\tfrac18(s-1)(s+2)\,\delta^2+\order{\delta^4}\Big],
\label{eq:a1}
\end{equation}
which lies $4\%$ below $s\delta$ for $s=2.71$ and $\delta=0.2$ (we use the exact value), and the ratio of the second harmonic to the square of the first is
\begin{equation}
\frac{a_2}{a_1^2}=\frac{s-1}{4s}
\label{eq:ratio}
\end{equation}
at leading order. In many constructions of black holes from inflation the enhancement is engineered in the potential~\cite{GangopadhyayPBH2022}, where loop corrections from sharp features can limit it~\cite{ChoudhuryNoGo2023}; here it is engineered in the friction, and the envelope $\mathcal A(k)$ follows from the coupled perturbations: a localized peak at the feature followed by a broad plateau across the window (Fig.~\ref{fig:spectrum}).

\subsection{The response kernel}
\label{sec:kernel}

The periodic friction is not static: it completes a cycle in about three e-folds, while a mode takes about half an e-fold to freeze. What matters then is how the power of a mode responds to a friction that changes during its freeze-out. A simple reduction shows what to expect. Deep in the window $\Upsilon$ is three to four times $3H$, and dropping $\delta\ddot\phi$ against the friction turns the inflaton equation into an overdamped Langevin equation~\cite{GrahamMoss2009,RamosdaSilva2013,BGRR2014fluid} with relaxation rate $\nu_k=(k^2/a^2+m_\phi^2)/(3H+\Upsilon)$. A mode tracks its local equilibrium while $\nu_k\gg H$ and freezes when $\nu_k\simeq H$, that is at
\begin{equation}
\frac{k}{a}\Big|_{\rm fo}\simeq\sqrt{3(1+Q)}\,H,
\label{eq:fo}
\end{equation}
about $1.3$ e-folds before horizon crossing for $Q\simeq3.4$. The width of freeze-out is set by how fast $\nu_k/H$ redshifts,
\begin{equation}
\Delta_N\equiv\Big|\frac{d\ln(\nu_k/H)}{dN}\Big|^{-1}=\Big|2-2\epsilon_H+\frac{d\ln(1+Q)}{dN}\Big|^{-1},
\label{eq:DeltaN}
\end{equation}
and is essentially independent of $Q$: the $\sqrt{1+Q}$ that fixes when a mode freezes drops out of how long it takes. In the window $\Delta_N=0.50$--$0.51$; at the onset of the feature it is not defined. With a geometrically redshifting rate, the frozen amplitude is an average of its attractor value against the window $W(N)=(\nu_k/H)\,e^{-\Delta_N\nu_k/H}$. The variance obeys $d\mathcal V/dN=-2\nu\mathcal V+D$ with a source $D\propto z^3S(\tilde z)$, so a modulation of the source by $e^{-i\omega N}$ reaches the frozen power filtered by
\begin{equation}
|\mathcal T_P(\omega)|=\frac{|\Gamma(\tfrac32+i\omega\Delta_N)|}{\Gamma(\tfrac32)}=\sqrt{\frac{1+4\omega^2\Delta_N^2}{\cosh(\pi\omega\Delta_N)}},
\label{eq:TP}
\end{equation}
while the amplitude would be filtered by $|\Gamma(1+i\omega\Delta_N)|$, which is smaller. The relevant parameter is $\omega\Delta_N\simeq1.9\times0.51\simeq0.97$, of order one, so the fundamental of the comb is suppressed by $|\mathcal T_P(\omega)|\simeq0.67$ and the second harmonic by $|\mathcal T_P(2\omega)|\simeq0.27$. A slower modulation is filtered less but fits fewer cycles into the window. The reduction neglects the variation of $S(\tilde z)$ across freeze-out (from $0.95$ to $0.87$ at $Q=3.28$), it filters a modulation of the source but not of the relaxation rate, and it assumes Markovian noise. For a noise correlation time $\tau_c=1/(\kappa T)$ the memory corrections are controlled by $(3H+\Upsilon)\tau_c=3(1+Q)/[\kappa(T/H)]$~\cite{GangopadhyayKumar2026}, about $0.5/\kappa$ in the window; in the coupled system they move the comb by at most $3.5\%$ for values of this parameter up to $4$ (Sec.~\ref{sec:coupled}).

\begin{figure*}[t]
\centering
\includegraphics[width=\textwidth]{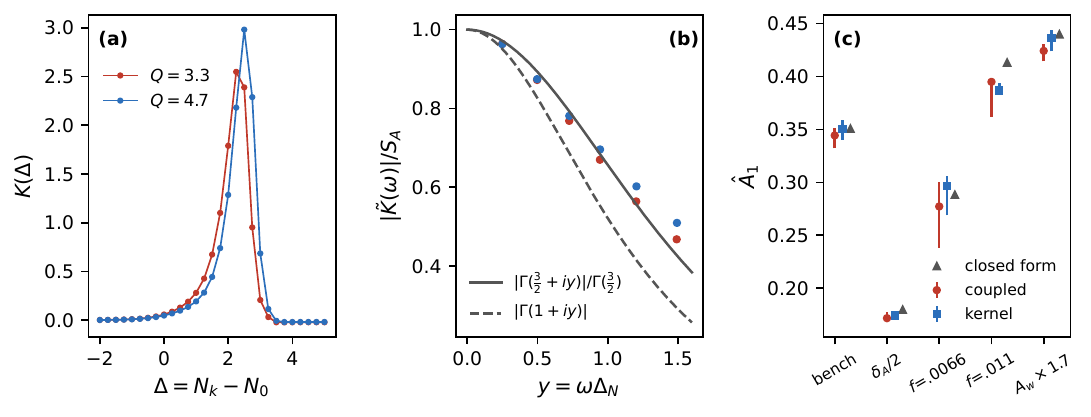}
\caption{The response of the coupled system. (a) The power-response kernel $K(\Delta)$, the change of $\ln\Pz$ per unit $\delta\ln A\,dN$ inserted $\Delta=N_k-N_0$ e-folds before the mode crosses the horizon at $N_k$, at $Q=3.3$ (red) and $Q=4.7$ (blue). (b) The modulus of its Fourier transform normalized by $S_A$ and corrected for the width of the insertion (points), against the reduced power filter $|\Gamma(3/2+iy)|/\Gamma(3/2)$ (solid) and amplitude filter $|\Gamma(1+iy)|$ (dashed), with $y=\omega\Delta_N$. (c) The comb fundamental for the five cases of Table~\ref{tab:sens}: coupled spectra (circles), kernel convolution (squares) and closed form (triangles), with bars for the spread over fit ranges.}
\label{fig:kernel}
\end{figure*}

We therefore measure the response of the full coupled system, in the spirit of the window functions of generalized slow roll in cold inflation~\cite{Stewart2002,DvorkinHu2010}, where the power responds to a change of the background through a known window in the time of crossing; here that window comes from friction-dominated freeze-out and has to be measured. A narrow multiplicative bump, $A\to A[1+\epsilon\,e^{-(\phi-\phi_0)^2/2\sigma_\phi^2}]$ with $\epsilon=0.04$ and a width of $\sigma_N=0.15$ e-fold, is inserted a time $\Delta$ before a fixed comoving mode crosses the horizon. The background is re-solved, the power of that mode is recomputed, and
\begin{equation}
K(\Delta)=\frac{\delta\ln\Pz}{\epsilon\sqrt{2\pi}\,\sigma_N}
\label{eq:kernel}
\end{equation}
is the response per unit $\delta\ln A\,dN$, with the change of the background included. We measured it on unmodulated window-only backgrounds at $Q=3.28$ ($T/H=25.4$) and $Q=4.74$ ($T/H=23.6$), for modes crossing $6.4$ and $9.5$ e-folds after their window opens, with $\Delta$ from $-2$ to $5$ in steps of $0.25$. The response is linear (halving $\epsilon$ changes it by less than $0.1\%$), and we divide its Fourier transform by the smoothing $e^{-\omega^2\sigma_N^2/2}$ of the insertion, which halving $\sigma_N$ confirms to $1\%$. The kernel integrates to the static response, $S_A=2.71$ and $2.88$. Its modulus, normalized by $S_A$, follows Eq.~\eqref{eq:TP} to within $3\%$ for $\omega\leq1.9$ at both points (Fig.~\ref{fig:kernel}); above that the kernel is larger, by $1$--$8\%$ at $\omega=2.4$ and $10$--$20\%$ at $\omega=3$. Its phase is not that of the reduction. The kernel is centred $2.0$ and $2.2$ e-folds before crossing, earlier than the $1.3$--$1.4$ of the reduced power window, and its rms width is $0.55$--$0.56$ against the reduced $0.48$, so at $\omega=1.9$ the two responses differ in phase by $1.6$--$1.7$~rad. One likely contributor is the periodic mass term $u\Upsilon_{,\phi}$ of Eq.~\eqref{eq:cdphi}, of order $4H^2$ in the window and comparable to $z^2\simeq13$ at freeze-out, which the reduction neglects against $k^2/a^2$. Equation~\eqref{eq:TP} is therefore a reduced-model approximation whose modulus the coupled system supports in the tested window, $3<Q<5$ and $\omega\lesssim2$. It is not an exact transfer function, and its phase should not be used.

\subsection{The comb}
\label{sec:comb}

The kernel turns the drive into a prediction with no free parameter: convolving it with $\delta\ln A=\ln[1+\delta_A\cos\theta]$ gives the modulated spectrum. The closed-form version is
\begin{equation}
\hat A_1=a_1(S_A,\delta_A)\,|\mathcal{T}_P(\omega)|.
\label{eq:A1hat}
\end{equation}
To compare either with the coupled spectrum we form the ratio $R=\mathcal P_{\rm mod}/\mathcal P_{\rm ref}$ of the modulated spectrum to that of the reference trajectory, interpolated to the same $k$, and fit it by linear least squares to
\begin{equation}
R=c_0+b_1x+b_2x^2+\sum_{n=1}^{3}\big[c_n\cos n\theta+s_n\sin n\theta\big].
\label{eq:fitmodel}
\end{equation}
Here $x$ is $\ln k$ normalized over the fitted modes, and $\theta=[\phi(N_k-\lambda)-\phi_c]/f$ is the coupling phase on the modulated trajectory a lag $\lambda$ before crossing, so that the template follows the drift of the frequency. The lag is chosen on a grid of $0.05$ by the smallest residual. The harmonics about the mean are $\hat A_n=\sqrt{c_n^2+s_n^2}/c_0$ with phases $\psi_n=\arg(c_n-is_n)$. The fit uses the modes crossing from two e-folds after the window opens, by which time the kernel no longer reaches back across the opening, to $0.3$ e-fold before it closes. The residuals are deterministic rather than noise, so we quote the spread of each estimate over four alternative fit ranges, with the start moved by $\pm0.5$ e-fold and the end by $\pm0.5$ e-fold.

\begin{table*}[t]
\caption{The comb fundamental $\hat A_1$ from the coupled calculation against three predictions. Each row is a pair of coupled solves, with and without modulation, on a window-only background, with modes every $0.05$ e-fold. $Q$ and $\omega$ are medians over the window and $\delta_Q$ is measured at fixed field; $\Delta_N=0.50$--$0.51$ in every row. Square brackets give the range over the alternative fit ranges. ``Kernel'' convolves the measured kernel ($Q=3.28$ for the first four rows, $4.74$ for the last) with the drive and applies the same estimator, and $\Delta$ is the deviation of the coupled value from it. ``Closed'' is Eq.~\eqref{eq:A1hat} with $S_A=2.71$ ($2.88$). ``$G$-based'' replaces $a_1(S_A,\delta_A)$ by $a_1(p,\delta_Q)$ with $p$ from Eq.~\eqref{eq:Q52} for the two fits of Eq.~\eqref{eq:GQ}.}
\label{tab:sens}
\footnotesize\setlength{\tabcolsep}{2pt}
\begin{tabular}{lcccccccc}
\hline\noalign{\smallskip}
Variant & $Q$ & $\omega$ & $\delta_Q$ & cycles & coupled & kernel & $\Delta$ & closed\,/\,$G$-based\\
\noalign{\smallskip}\hline\noalign{\smallskip}
benchmark window & $3.39$ & $1.90$ & $0.161$ & $2.1$ & $0.344$ [$0.332$--$0.351$] & $0.351$ [$0.340$--$0.359$] & $-1.8\%$ [$-2.6$, $-1.2$] & $0.351$\,/\,$0.450$--$0.465$\\
$\delta_A=0.1$ & $3.30$ & $1.93$ & $0.080$ & $2.1$ & $0.172$ [$0.168$--$0.177$] & $0.174$ [$0.170$--$0.179$] & $-1.5\%$ [$-2.4$, $-0.9$] & $0.180$\,/\,$0.231$--$0.239$\\
$f=0.0066$ & $3.32$ & $2.42$ & $0.169$ & $2.6$ & $0.277$ [$0.238$--$0.301$] & $0.297$ [$0.269$--$0.306$] & $-6.5\%$ [$-11.4$, $-1.6$] & $0.288$\,/\,$0.383$--$0.396$\\
$f=0.011$ & $3.25$ & $1.46$ & $0.155$ & $1.6$ & $0.395$ [$0.362$--$0.398$] & $0.387$ [$0.385$--$0.394$] & $+2.1\%$ [$-8.2$, $+2.5$] & $0.413$\,/\,$0.506$--$0.522$\\
$A_w\times1.7$ & $4.82$ & $1.43$ & $0.145$ & $2.5$ & $0.424$ [$0.414$--$0.431$] & $0.436$ [$0.424$--$0.444$] & $-2.8\%$ [$-3.1$, $-1.9$] & $0.440$\,/\,$0.523$--$0.555$\\
\noalign{\smallskip}\hline
\end{tabular}
\end{table*}

Table~\ref{tab:sens} tests the predictions on the benchmark window and on four variations that change the depth, the frequency and the strength of the drive, each on a window-only background so that the comb is seen by itself. The kernel convolution lies $1.5$--$2.8\%$ above the coupled fundamental in three rows, $6.5\%$ above it for the fastest modulation, where the spread over fit ranges is also largest, and $2\%$ below it for the slowest, which keeps only $1.6$ cycles in the fit. The closed form lies $2$--$5\%$ above the coupled values. The $G$-based estimate is $23$--$43\%$ high, because the exponent of the fits, obtained for another placement of the noise, lies above the measured static response and because the dynamic $\delta_Q$ already contains part of the time dependence. The fundamental scales as it should: halving $\delta_A$ halves it, a faster modulation is suppressed by the filter, and a stronger window raises it through $S_A$. Because the kernel is measured with the same solver, this agreement shows that the comb is the linear response of the coupled system to the drive, with a response that depends only on the time difference; it is not an independent calculation of the comb. The independent content is the closed form, which matches the coupled response in modulus only.

For the full benchmark, feature included, the same estimator gives $\hat A_1=0.342$ over $2.3$ cycles ($0.333$--$0.350$ across fit ranges, on the grid of $0.2$ e-fold used for the figures; Table~\ref{tab:closure} uses modes every $0.05$ e-fold), with $\lambda=2.15$ and $\psi_1=0.063$ (Fig.~\ref{fig:spectrum}). This is within one percent of the window-only value, although the two windows open very differently. The best-fit lag agrees with the centroid of the kernel, not with that of the reduced window. The residual has an rms of $0.026$ of the mean, about $8\%$ of $\hat A_1$, and it is structure from the drifting frequency rather than noise. The second harmonic is less robust. With little more than two cycles and a frequency drifting between $1.7$ and $2.2$, the extracted $\hat A_2/\hat A_1^2$ ranges from $0.16$ to $0.28$ with the fit range, while filtering Eq.~\eqref{eq:ratio} with Eq.~\eqref{eq:TP} gives $0.11$; the coupled data do not resolve the ratio, and we make no claim for it.

The comb is a ratio and a response, and it is much less sensitive to the noise closure than the power itself. We recomputed the window modes of the benchmark under each alternative closure, with and without modulation, and remeasured the kernel (Table~\ref{tab:closure}). The window power moves by factors between $0.77$ and $1.29$, but $\hat A_1$ changes by at most $3.9\%$, its phase by at most $0.037$~rad, $S_A$ by at most $3.9\%$, and the filtered modulus at $\omega=1.9$ by $0.2\%$. This holds where the window opens smoothly. On the window-only background, where it opens abruptly, white noise makes the modes inside the horizon at the opening ring: the power of modes crossing $2.5$ e-folds after the opening is about $22$ times the later plateau, and the ringing lasts until about $3.8$ e-folds in. The fit then gives $0.385$ with white noise against $0.344$ with the occupation weighting, $12\%$ higher, while a fit starting four e-folds in, after the ringing, gives $0.348$ and $0.337$. The occupation weighting suppresses the noise inside the horizon and with it the ringing. The closure dependence of the comb is therefore a few percent on the benchmark, and up to $12\%$ after an abrupt opening with white noise.

\begin{table*}[t]
\caption{The comb under alternative noise closures. For each closure the window modes of the benchmark are recomputed with and without modulation, every $0.05$ e-fold, and the kernel at $Q=3.28$ is remeasured. ``Window power'' is the unmodulated power of a mode crossing five e-folds after the window opens, relative to the benchmark. $\Delta\psi_1$ is the change of the phase of the fundamental at the benchmark lag, and $\lambda$ the best-fit lag. The last three columns describe the kernel: its centroid (e-folds before crossing), its deconvolved rms width and its Fourier modulus at $\omega=1.9$ normalized by $S_A$.}
\label{tab:closure}
\footnotesize\setlength{\tabcolsep}{3pt}
\begin{tabular}{lcccccccc}
\hline\noalign{\smallskip}
Closure & window power & $\hat A_1$ & $\Delta\psi_1$ [rad] & $\lambda$ & $S_A$ & centroid & width & $|\tilde K(1.9)|/S_A$\\
\noalign{\smallskip}\hline\noalign{\smallskip}
$S(\tilde z)$ (benchmark) & $1.00$ & $0.342$ [$0.331$--$0.349$] & $+0.000$ & $2.15$ & $2.71$ & $2.03$ & $0.548$ & $0.670$\\
white noise & $1.29$ & $0.354$ [$0.342$--$0.361$] & $+0.030$ & $2.15$ & $2.80$ & $2.05$ & $0.550$ & $0.670$\\
Gaussian cutoff & $1.13$ & $0.344$ [$0.333$--$0.351$] & $+0.003$ & $2.15$ & $2.72$ & $2.03$ & $0.547$ & $0.671$\\
$S(\tilde z/2)$ & $1.14$ & $0.348$ [$0.336$--$0.355$] & $+0.016$ & $2.15$ & $2.76$ & $2.04$ & $0.549$ & $0.670$\\
$S(2\tilde z)$ & $0.77$ & $0.329$ [$0.318$--$0.336$] & $-0.037$ & $2.15$ & $2.60$ & $2.01$ & $0.546$ & $0.670$\\
\noalign{\smallskip}\hline
\end{tabular}
\end{table*}

\section{Primordial black holes}
\label{sec:pbh}

When a mode of the feature re-enters the horizon during radiation domination, a rare overdense region can collapse to a black hole. We compute the formation fraction from the compaction function~\cite{Musco2019,YoungByrnesSasaki2014,HaradaYooKohri2013} rather than with the Press--Schechter prescription~\cite{PressSchechter1974,CarrHawking1974,Carr1975}, with a variance, a profile and a threshold that refer to the same random variable~\cite{Musco2021,Ferrante2023}. Everything in this section is a conditional Gaussian estimate: it assumes Gaussian curvature perturbations and the thermal background of Sec.~\ref{sec:background}, whose onset is not under control.

The linear compaction is $C_l(r)=-\tfrac43\,r\,\zeta'(r)$ and the full one $C=C_l-\tfrac38C_l^2$. Its peak, at the scale $r_m$, and the shape parameter $\alpha$ of the collapsing profile follow from the spectrum,
\begin{equation}
\int d\ln k\,\Big[(k^2r_m^2-1)\frac{\sin kr_m}{kr_m}+\cos kr_m\Big]\Pz(k)=0,
\label{eq:rmeq}
\end{equation}
\begin{equation}
\begin{aligned}
F(\alpha)\big[1+F(\alpha)\big]\alpha&=-\tfrac12\big(1+r_m\mathcal J_c/\mathcal J_s\big),\\
F(\alpha)&=\sqrt{1-\tfrac25\,e^{-1/\alpha}\,\alpha^{1-5/2\alpha}/\gamma_\alpha},
\end{aligned}
\label{eq:alpha}
\end{equation}
with $\mathcal J_c=\int d\ln k\,k^2\cos(kr_m)\Pz$, $\mathcal J_s=\int d\ln k\,k\sin(kr_m)\Pz$ and $\gamma_\alpha=\Gamma(\tfrac5{2\alpha})-\Gamma(\tfrac5{2\alpha},\tfrac1\alpha)$. The threshold is the fit to numerical collapse of Ref.~\cite{Musco2021},
\begin{equation}
C_c(\alpha)=\frac{4}{15}\,e^{-1/\alpha}\,\frac{\alpha^{1-5/2\alpha}}{\gamma_\alpha},
\label{eq:Cc}
\end{equation}
and the variance of the linear compaction uses the same top-hat smoothing and the radiation-era transfer function at horizon crossing~\cite{Ferrante2023},
\begin{equation}
\sigma_c^2=\frac{16}{81}\int d\ln k\,(kr_m)^4\,W^2(kr_m)\,\mathcal{T}_r^2(kr_m)\,\Pz(k),
\label{eq:sigmac}
\end{equation}
with $W(x)=3(\sin x-x\cos x)/x^3$ and $\mathcal T_r(x)=W(x/\sqrt3)$. A region collapses when $C_l$ exceeds $C_{l,c}=\tfrac43\big(1-\sqrt{1-\tfrac32C_c}\big)$, and the mass follows the critical scaling of numerical collapse~\cite{Choptuik1993,NiemeyerJedamzik1999,ShibataSasaki1999}, $M=\mathcal{K}M_H(C-C_c)^{\gamma}$, with $\gamma\simeq0.36$ for radiation~\cite{EvansColeman1994,MuscoMiller2013} and $\mathcal{K}\simeq4$, a profile-dependent constant~\cite{NiemeyerJedamzik1999,MuscoMiller2013}. Taking formation at the single scale $r_m$, the mass fraction per logarithmic mass interval is
\begin{equation}
\frac{d\beta}{d\ln M}=\frac{M}{M_H}\,\frac{e^{-C_l^2/2\sigma_c^2}}{\sqrt{2\pi}\,\sigma_c}\,\frac{C-C_c}{\gamma\,(1-\tfrac34C_l)},
\label{eq:beta}
\end{equation}
and the present abundance relative to the dark matter is
\begin{equation}
\frac{d\fpbh}{d\ln M}=\frac{\Omega_rh^2}{\Omega_{\rm DM}h^2}\,\frac{g_*}{g_{*0}}\Big(\frac{g_{*s0}}{g_{*s}}\Big)^{4/3}\frac{a_0}{a_f}\,\frac{d\beta}{d\ln M},
\label{eq:FDM}
\end{equation}
with $a_f$ the scale factor at re-entry of $1/r_m$, $\Omega_{\rm DM}h^2=0.12$, and $M_H(k)=14.0\,\Msun\,(k/10^6\Mpc^{-1})^{-2}$ for $g_*=106.75$~\cite{GreenKavanagh2021,CarrKohriSendouda2021}. Accretion, mergers and evaporation are neglected.

\begin{figure}[t]
\centering
\includegraphics[width=\columnwidth]{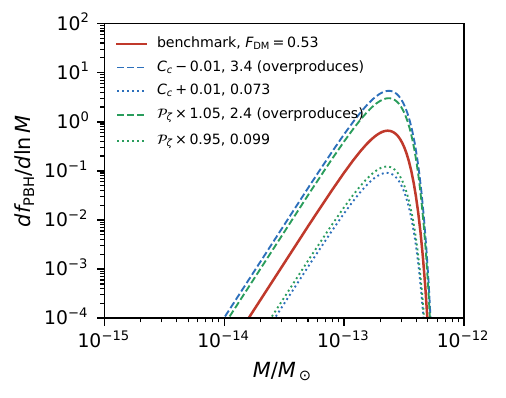}
\caption{The black-hole mass function of the benchmark, a conditional Gaussian estimate (red, $F_{\rm DM}=0.53$). Blue: the threshold lowered (dashed) or raised (dotted) by $0.01$. Green: the curvature power scaled by $1.05$ (dashed) or $0.95$ (dotted). Curves with $F_{\rm DM}>1$ would overproduce dark matter and only show how sensitive the abundance is.}
\label{fig:pbh}
\end{figure}

For the benchmark the profile is computed from the feature alone, between $\ln\kpk-5$ and the spectral minimum $2.1$ above $\ln\kpk$ that separates it from the window. We find $\kpk r_m=1.85$, $\alpha=0.883$, $C_c=0.4828$ and $C_{l,c}=0.633$, and $\sigma_c=0.0803$, so that $C_{l,c}/\sigma_c=7.89$. As a check, a narrow peak gives $kr_m=2.74$ and $C_c=0.595$, the values of Ref.~\cite{Musco2021}. Formation occurs at $M_H=3.44\times10^{-13}\Msun$ and $T\simeq3.7\times10^5$~GeV, with $\beta=8.3\times10^{-16}$ and a dark-matter fraction
\begin{equation}
F_{\rm DM}\equiv\int\frac{d\fpbh}{d\ln M}\,d\ln M\simeq0.53.
\label{eq:FDMval}
\end{equation}
The mass function peaks at $2.3\times10^{-13}\Msun$, with $90\%$ of it between $1.0$ and $3.3\times10^{-13}\Msun$ (Fig.~\ref{fig:pbh}). These masses lie in the asteroid-mass window, where black holes can still be all of the dark matter, above the evaporation bounds and below the masses probed by microlensing~\cite{Niikura2019HSC,Niikura2019OGLE,CarrKuhnel2020}; compilations of the constraints are given in Refs.~\cite{GreenKavanagh2021,CarrKohriSendouda2021,CarrKuhnel2020}, and Ref.~\cite{CarrKuhnelVisinelli2021} discusses the opposite end of the mass range, black holes far heavier than these. The mass function is narrow because the feature is a single resonant crossing, and the window, two orders of magnitude lower, never seeds a second population. We have not applied extended-mass constraints, since the abundance is an illustration.

The abundance is not a prediction. We chose $A_b$ so that $F_{\rm DM}$ is of order one, and because $\beta$ is exponential in the amplitude, a few percent less gives orders of magnitude fewer holes. Table~\ref{tab:budget} lists the sensitivities, each from a recomputation. Lowering $A_b$ by $6.7\%$ lowers the peak power to $5.3\times10^{-3}$ and $F_{\rm DM}$ to $9\times10^{-6}$. Scaling the power by $\pm5\%$ moves $F_{\rm DM}$ by a factor of five, and a shift of $0.01$ in the threshold by a factor of seven. The noise closures of Sec.~\ref{sec:coupled} span $F_{\rm DM}=10^{-10}$--$10^{5}$. The steepness is generic to rare, near-Gaussian threshold collapse~\cite{YoungByrnesSasaki2014,CarrKuhnel2020,SasakiSuyama2018,CarrKuhnelSandstad2016,CarrKohriSendouda2021}: $\partial\ln\beta/\partial\ln\sigma_c\simeq(C_{l,c}/\sigma_c)^2\simeq62$, so a one-percent change in $\sigma_c$ moves $\beta$ by a factor of about two, while our linear amplitude is uncertain at the level of tens of percent through the closure alone. Matching the variance to the threshold matters as well: a Gaussian-smoothed density variance paired with the same threshold would give $F_{\rm DM}\sim10^{-167}$. The mass scale is steadier. Across all the recomputed spectra the peak of the mass function stays between $1.9$ and $2.8\times10^{-13}\Msun$ and the threshold between $0.476$ and $0.488$.

\begin{table}[t]
\caption{Sensitivity of the black-hole abundance. Each entry is a recomputation with the matched prescription. Rows marked ``spectrum'' recompute the coupled spectrum on the coarser grid of the noise runs, on which the benchmark gives $F_{\rm DM}=0.41$. In those rows the variance omits the window; on the same grid, including it gives $1.07$, which shows how sensitive the abundance is to $\sigma_c$. The entries are sensitivities about a tuned benchmark, not error bars on a prediction.}
\label{tab:budget}
\small\setlength{\tabcolsep}{3pt}
\begin{tabular}{lcc}
\hline\noalign{\smallskip}
Change & $F_{\rm DM}$ & $M_{\rm pk}/10^{-13}\Msun$ \\
\noalign{\smallskip}\hline\noalign{\smallskip}
benchmark & $0.53$ & $2.3$\\
$\mathcal{K}=3.3$; $4.4$ & $0.44$; $0.58$ & $1.9$; $2.5$\\
$\gamma=0.33$; $0.38$ & $0.62$; $0.48$ & $2.7$; $2.1$\\
$C_c-0.01$; $C_c+0.01$ & $3.4$; $0.073$ & $2.4$; $2.3$\\
profile up to $k_{\rm min}e^{-0.5}$ & $2.3$ & $2.4$\\
no transfer in $\sigma_c$ & $3\times10^5$ & $2.8$\\
$\Pz\times0.95$; $\times1.05$ & $0.099$; $2.4$ & $2.3$; $2.4$\\
sampling $\Delta N_k=0.1$; $0.4$ & $0.54$; $0.39$ & $2.3$; $2.3$\\
$A_b\times0.933$ (spectrum) & $9\times10^{-6}$ & $1.9$\\
$A_b\times1.005$ (spectrum) & $1.05$ & $2.4$\\
Gaussian cutoff (spectrum) & $4.2$ & $2.5$\\
$S(\tilde z/2)$ (spectrum) & $4.4\times10^2$ & $2.5$\\
$S(2\tilde z)$ (spectrum) & $1.4\times10^{-10}$ & $2.1$\\
white noise (spectrum) & $8.5\times10^4$ & $2.7$\\
\noalign{\smallskip}\hline
\end{tabular}
\end{table}

Two corrections are not included. Peaks theory~\cite{YoungByrnesSasaki2014,YooHaradaGarriga2018} changes $\beta$ at fixed amplitude, and non-Gaussianity matters more. The tree-level bispectrum of Sec.~\ref{sec:bispectrum} gives $f_{\rm NL}\simeq-3.8$ in the equilateral configuration at the peak. The threshold of Ref.~\cite{Musco2021} contains the non-Gaussianity of the relation between $\zeta$ and the compaction, but not intrinsic non-Gaussianity of $\zeta$, and for a threshold almost eight standard deviations above the mean an $f_{\rm NL}$ of order one could move the abundance by orders of magnitude in either direction. A controlled abundance needs the distribution of the compaction itself~\cite{AtalGermani2019,Ferrante2023,Pi2024}, on a bath shown to follow the onset of the feature. We therefore present $F_{\rm DM}\simeq0.5$ as an illustrative Gaussian benchmark.

\section{Induced gravitational waves}
\label{sec:sigw}

The enhanced curvature power also sources tensor modes at second order~\cite{AnandaClarksonWands2007,BaumannSteinhardt2007,SaitoYokoyama2009}, a standard probe of small-scale power and of the black-hole hypothesis~\cite{BartoloDeLuca2019,CapriniFigueroa2018}. We compute the radiation-era part for a Gaussian source; tensor modes sourced during inflation or by the bath are not included (Ref.~\cite{Ballesteros2022} discusses some of them for a transient dissipative peak). Today's spectral density is~\cite{Domenech2021}
\begin{equation}
h^2\Ogw(k)=1.61\times10^{-5}\left(\frac{g_*}{106.75}\right)^{\!-1/3}\!\mathcal{I}(k),
\label{eq:OmGW}
\end{equation}
\begin{equation}
\begin{aligned}
\mathcal{I}(k)=\frac16\int_0^\infty\! dv&\int_{|1-v|}^{1+v}\! du\;
\left[\frac{4v^2-(1+v^2-u^2)^2}{4uv}\right]^2\\
&\times\overline{I^2(u,v)}\;\Pz(uk)\,\Pz(vk),
\end{aligned}
\label{eq:Ik}
\end{equation}
with the time-averaged radiation-era kernel~\cite{KohriTerada2018,EspinosaRaccoRiotto2018}
\begin{equation}
\begin{aligned}
\overline{I^2(u,v)}&=\frac12\left[\frac{3s_{uv}}{4u^3v^3}\right]^2
\Big\{\big[-4uv+s_{uv}\,\ell_{uv}\big]^2\\
&\qquad+\pi^2s_{uv}^2\,\Theta\big(u+v-\sqrt3\big)\Big\},\\
s_{uv}&=u^2+v^2-3,\qquad
\ell_{uv}=\ln\Big|\frac{3-(u+v)^2}{3-(u-v)^2}\Big|.
\end{aligned}
\label{eq:Ibar}
\end{equation}
The factor $1/6$ comes from $\Ogw=(k/aH)^2\overline{\mathcal P_h}/24$ with a factor $4$ in the tensor power. For a flat spectrum $\Pz=A$ the double integral without it is $4.933\,A^2$, and $\mathcal I=0.8222\,A^2$, the value of Ref.~\cite{KohriTerada2018}, which checks the normalization. On the computed spectrum the relative quadrature error is $2\times10^{-8}$ at the peak.

\begin{figure*}[t]
\centering
\includegraphics[width=\textwidth]{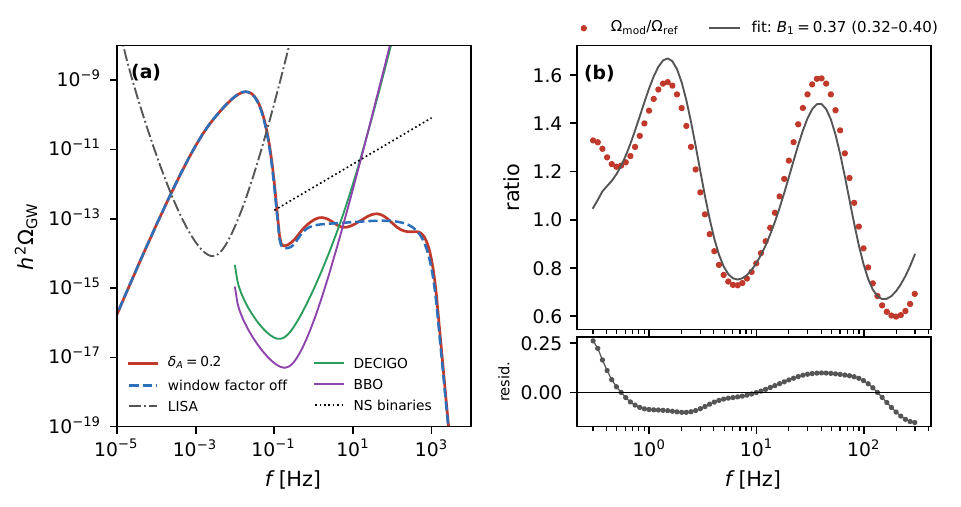}
\caption{The induced gravitational waves for a Gaussian source. (a) The spectrum with the periodic factor of the window on (red) and off (blue dashed), against the power-law-integrated sensitivities of LISA ($T=4$~yr), DECIGO and BBO ($T=3$~yr, idealized co-located pairs), and the neutron-star binary foreground before subtraction (dotted). The millihertz peak comes from the feature and is conditional; the window gives the band from $0.2$~Hz to $0.9$~kHz. (b) The ratio of the two spectra over $0.28$--$300$~Hz with the fit at the scalar coupling phase, $B_1=0.37$, and below it the residual relative to the mean.}
\label{fig:gw}
\end{figure*}

The feature gives a peak $h^2\Ogw\simeq4.6\times10^{-10}$ at $f_{\rm pk}\simeq19$~mHz, with $f=1.546\times10^{-15}(k/\Mpc^{-1})$~Hz, in the LISA band~\cite{LISA2017,LISA2024} (Fig.~\ref{fig:gw}). The power-law-integrated sensitivity~\cite{ThraneRomano2013} for the LISA noise model of Ref.~\cite{RobsonCornishLiu2019}, with four years and a threshold of one, is $3.9\times10^{-13}$ there, three orders of magnitude below the peak. The peak sits at $1.04\,\kpk$, and its height is not tied to the abundance: across the recomputed spectra of Table~\ref{tab:budget} it follows $(\Pz^{\rm pk})^{1.99}$, ranging over $1.6\times10^{-10}$--$1.2\times10^{-9}$, while $F_{\rm DM}$ ranges over fifteen orders of magnitude. A detection would measure the peak power, not the dark-matter fraction. It is conditional on Gaussian statistics, since a non-Gaussian source adds connected four-point contributions~\cite{AdsheadLozanovWeiner2021,CaiPiSasaki2019} that we have not computed. Below the peak the spectrum approaches the causal slope $f^3$ with its logarithmic running~\cite{YuanChenHuang2020}. With formation at $1/r_m=\kpk/1.85$, the black-hole mass and the peak frequency are related by $M\simeq2.3\times10^{-13}\Msun$ for $f_{\rm pk}=19$~mHz and $M\propto f_{\rm pk}^{-2}$ at fixed peak shape, a relation that holds for any radiation-era collapse with these peak-shift factors.

The window sources a second band. Its curvature plateau, $\Pz\simeq(0.5$--$1.1)\times10^{-4}$ for $k\simeq3\times10^{14}$--$4\times10^{17}\Mpc^{-1}$, induces gravitational waves from $0.2$~Hz to $0.9$~kHz at $h^2\Ogw\simeq10^{-14}$--$1.4\times10^{-13}$ ($9\times10^{-14}$ at $1$~Hz). This is within the reach of DECIGO~\cite{Kawamura2011} and BBO~\cite{YagiSeto2011}, far above the PTA band~\cite{NANOGrav2023}, and its upper end overlaps the band of third-generation ground-based detectors~\cite{Punturo2010,Maggiore2020}, which we do not forecast. The total, $\int h^2\Ogw\,d\ln f\simeq9\times10^{-10}$, is dominated by the feature and well inside the BBN bound~\cite{CapriniFigueroa2018}.

This band carries the comb, though not unchanged, because the induced spectrum is a convolution of two copies of $\Pz$. The quadratic source mixes the fundamental with its second harmonic and averages both over the convolution, as worked out for resonant features in Ref.~\cite{Fumagalli2021} and for different expansion histories in Ref.~\cite{Witkowski2022}. The structure follows semi-analytically if the phases are kept. Writing $\Pz(uk)=\mathcal{A}(uk)[1+m(uk)]$ with $m={\rm Re}\sum_n a_n e^{in\theta(k)}e^{-in\omega\ln u}$ and $a_n=\hat A_ne^{i\psi_n}$, since $d\theta/d\ln k\simeq-\omega$, the fractional change of the induced spectrum is
\begin{equation}
\frac{\Ogw^{\rm mod}}{\Ogw^{\rm ref}}-1=\left\langle\,m(uk)+m(vk)+m(uk)m(vk)\,\right\rangle_w,
\label{eq:residual}
\end{equation}
an average over the $(u,v)$ plane weighted by the integrand of Eq.~\eqref{eq:Ik} with the unmodulated envelope. In terms of the moments $M_n=\langle\tfrac12(e^{-in\omega\ln u}+e^{-in\omega\ln v})\rangle_w$ and $M_{n,m}=\langle e^{-i\omega(n\ln u+m\ln v)}\rangle_w$, to second order in the amplitudes,
\begin{equation}
\begin{aligned}
B_1e^{i\beta_1}&\simeq2a_1M_1,\qquad
B_2e^{i\beta_2}\simeq2a_2M_2+\tfrac12a_1^2M_{1,1},\\
\bar r&\simeq\tfrac12|a_1|^2\,{\rm Re}\,M_{1,-1}.
\end{aligned}
\label{eq:Bsemi}
\end{equation}
The second harmonic inherits a piece $\propto a_1^2$ from the self-convolution of the fundamental, and the product shifts the mean. At the centre of the band, with $\omega=1.89$, the moments are $M_1=0.574\,e^{0.03i}$, $M_2=0.378\,e^{0.57i}$, $M_{1,1}=0.628\,e^{0.65i}$ and $M_{1,-1}=0.705$, which give $B_1=0.39$ with $\beta_1-\psi_1=0.03$~rad, $B_2=0.059$ and a mean shift of $0.04$. A direct fit of the computed ratio over $0.28$--$300$~Hz, with Eq.~\eqref{eq:fitmodel} at the scalar coupling phase, gives $B_1=0.37$ ($0.32$--$0.40$ as the band edges move by half an e-fold), $\beta_1-\psi_1=0.05$~rad and $B_2=0.064$, in agreement with the moments. The residual has an rms of $0.09$ of the mean, largest near the edges of the band, where a fixed-frequency template cannot follow the convolution of a drifting comb. The comb therefore survives the quadratic source with $30$--$40\%$ contrast. It lives in this band only: the feature completes less than one cycle of the coupling, so the millihertz peak carries no comb.

Resolving an oscillation needs far more signal than detecting a background, so we forecast the comb with its own statistic. We use the Yagi--Seto noise spectra of DECIGO and BBO~\cite{YagiSeto2011}, multiplied by five for the sky average of a $90^\circ$ interferometer (a $60^\circ$ triangle would raise the noise by a further $4/3$) and converted as $\Omega_n=2\pi^2f^3S_{\rm eff}/3H_0^2$, for an idealized co-located and co-aligned pair observing for three years. The signal-to-noise ratio is ${\rm SNR}^2=2T\int df\,[s(f)/\sigma(f)]^2$, where $\sigma^2=(\Omega_n+\Omega_c)^2+\Omega_c^2$ is the variance of the cross-correlation and $\Omega_c=\Ogw+\epsilon_{\rm fg}\Omega_{\rm NS}$ contains the signal's own variance and a residual fraction $\epsilon_{\rm fg}$ of the neutron-star binary foreground, $h^2\Omega_{\rm NS}\simeq4\times10^{-13}$ at $0.3$~Hz and $8\times10^{-13}$ at $1$~Hz before subtraction~\cite{Yagi2013}. The power-law-integrated curves reach $h^2\Ogw=3.4\times10^{-17}$ at $0.13$~Hz (DECIGO) and $5.0\times10^{-18}$ at $0.18$~Hz (BBO), in line with Ref.~\cite{Schmitz2020}. The whole background is detected at ${\rm SNR}\simeq3\times10^3$ and $4\times10^3$, and the window band alone ($0.2$--$100$~Hz) at $266$ and $2.2\times10^3$.

For the comb, the signal is the difference $\Omega_{\rm mod}-\Omega_{\rm ref}$ over $0.2$--$100$~Hz, made orthogonal in the Fisher metric to everything a smooth null model can absorb. Without nuisance parameters its signal-to-noise ratio would be $57$ (DECIGO) and $488$ (BBO). Marginalizing the amplitude of the unmodulated template lowers these to $10$ and $84$. Adding quadratic distortions of the template in $\ln f$, together with the amplitude and slope of the foreground, lowers them to $2.4$ and $16$, and cubic distortions to $1.0$ and $8.1$. If the frequency and phase of the comb are not known in advance and are marginalized too, the numbers become $1.7$ and $12$ with quadratic freedom and $0.7$ and $6.1$ with cubic freedom. Leaving $10\%$ of the foreground unsubtracted gives $1.0$ and $7.2$ with cubic freedom, and leaving all of it $0.9$ and $3.7$. In this idealized forecast a BBO-class detector therefore reaches a signal-to-noise ratio of 6--16 on the comb if the binary foreground is subtracted completely, 5--14 if $10\%$ of it is left, and 2--8 if none of it is subtracted, depending on the freedom given to the null model, while the adopted DECIGO configuration does not resolve it. A realistic response, overlap function and foreground model would lower these numbers, so they are not detection significances.

\section{The bispectrum}
\label{sec:bispectrum}

Warm inflation produces a non-Gaussianity whose sign, size and shape are set by the dissipation~\cite{MossXiong2007,MossYeomans2011}. For a temperature-dependent friction on a slowly varying background, the second-order calculation of Ref.~\cite{BGMR2014NG} finds $|f_{\rm NL}|\simeq10$ in the strong regime, and that of Ref.~\cite{MirbabayiGruzinov2023} an equilateral amplitude that tends to $4.5$; both find shapes that overlap little with the standard templates. Our CMB pivot is effectively cold, so none of this reaches the scales that Planck constrains~\cite{Planck2018IX}. On small scales the background varies quickly and the friction is modulated, so the bispectrum has to be computed for this background.

The squeezed limit is fixed in general, provided its premise holds. If the long mode $\zeta_L$ has become adiabatic when the short modes freeze, its effect on them is a shift of the local scale factor,
\begin{equation}
\Pz(k)\big|_{\zeta_L}=\Pz\!\big(k\,e^{-\zeta_L}\big)\simeq\Pz(k)\Big[1-\zeta_L\,\frac{d\ln\Pz}{d\ln k}\Big],
\label{eq:modP}
\end{equation}
and in the normalization in which the single-field relation reads $f_{\rm NL}=\tfrac5{12}(1-n_s)$~\cite{Maldacena2003,CreminelliZaldarriaga2004},
\begin{equation}
f_{\rm NL}^{\rm sq}(k_s)=-\frac{5}{12}\,\frac{d\ln\Pz}{d\ln k_s}.
\label{eq:fsq}
\end{equation}
For the comb, $d\ln\Pz/d\ln k\simeq\omega\hat A_1\sin(\theta+\psi_1)$ plus the slope of the envelope, so the squeezed $f_{\rm NL}$ oscillates with amplitude $\tfrac5{12}\omega\hat A_1\simeq0.27$. We write the oscillating part of $f_{\rm NL}$ as $F_1\cos(\theta+\psi_1+\Delta\psi)$, so that $\Delta\psi$ is its phase offset from the power comb at the same lag. Because $\theta$ decreases with $\ln k$, a positive $\Delta\psi$ places the maxima of $f_{\rm NL}$ at larger $k$ than those of the power, by $\Delta\psi/\omega$ in $\ln k$. The squeezed relation gives $\Delta\psi=\pi/2$. The premise can be checked, because the solver evolves the radiation and the metric together with the inflaton. We followed long modes that leave the horizon at the pivot, at the onset of the feature, and $20$, $10$ and $2$ e-folds before the window opens. While the window modes freeze, the curvature power of these long modes changes by at most $3.2\times10^{-5}$, and two measures of a non-adiabatic part (the difference between the comoving curvatures of the inflaton and of the radiation, and the radiation density contrast on comoving slices) are at most $6.5\times10^{-7}$ of $\langle\mathcal R^2\rangle$ in power. Outside the horizon the radiation moves with the inflaton even where the dissipation is strong, so Eq.~\eqref{eq:modP} applies to the comb. It does not apply as cleanly to the feature: a mode four e-folds outside the horizon when the feature switches on changes by $4\%$ through the onset, although through the window it is constant to $10^{-7}$. Two further points limit what the relation says. It fixes only the squeezed configuration, and its leading term is removed by a change of coordinates for a local observer~\cite{TanakaUrakawa2011,PajerSchmidtZaldarriaga2013}, so it is a consistency condition on the calculation rather than an observable. For features in cold inflation the oscillatory bispectrum is tied to the power spectrum beyond the squeezed limit~\cite{FlaugerPajer2011,BehbahaniGreen2012,AdsheadDvorkin2012,ChenEastherLim2008,ChenFeatures2012}, but for a modulated friction no such relation is known, and the other configurations have to be computed.

\subsection{Second-order equations}
\label{sec:second}

We compute the other configurations at tree level by expanding the equations of the inflaton and of the radiation fluid to second order about the integrated background, with the metric left unperturbed as in Refs.~\cite{BGMR2014NG,MirbabayiGruzinov2023}. Time is $N$, a prime is $d/dN$, and a mode is labelled by $\bm\kappa=\bm k/aH$ with $z=|\bm\kappa|$. The first-order variables are those of Eq.~\eqref{eq:cov} without $\Psi$, $X=(\delta\phi,\delta\pi,e,\hat q)$ with $\pi=\phi'$, $e=\delta E/H^2$ for the radiation energy density $E=T^{00}$, and $\hat q=q_r/H$. They obey $X'=\mathsf AX+\hat n\hat\xi$, with $\mathsf A$ the drift of Eqs.~\eqref{eq:cdphi}--\eqref{eq:cdq} at $\Psi=0$, $\hat n=(0,1,-v,0)$, and a noise with $\langle\hat\xi\hat\xi\rangle'=\sigma^2\delta(N-N')$ and $\sigma^2=6Q(T/H)z^3S(\tilde z)H^2$. At second order $X^{(2)\prime}=\mathsf AX^{(2)}+\mathcal S_V+\mathcal S_\xi$, where $\mathcal S_V$ is quadratic in the first-order fields and $\mathcal S_\xi$ is bilinear in the noise and a field; a quadratic term carries two momenta with $\bm\kappa_1+\bm\kappa_2=\bm\kappa$.

The friction enters covariantly, as $\Upsilon u^\mu\partial_\mu\phi$ with $\Upsilon=A(\phi)T^3$, $u^\mu$ the velocity of the radiation in its energy frame and $T$ its rest-frame temperature. With $\hat\rho=\rho_r/H^2$, $\ell_1=d\ln A/d\phi$, $\ell_2=d^2\ln A/d\phi^2$ and the fluid three-velocity $v_r^i=\partial_iq_r/(\tfrac43\rho_r)$, its response to second order is $\delta\Upsilon/\Upsilon=\mathcal U_1+\mathcal U_2$ with
\begin{equation}
\begin{aligned}
\mathcal U_1&=\ell_1\delta\phi+\tfrac34\,e/\hat\rho,\\
\mathcal U_2&=\tfrac12\big(\ell_2+\ell_1^2\big)\delta\phi^2+\tfrac34\,\ell_1\delta\phi\,e/\hat\rho-\tfrac{3}{32}\,e^2/\hat\rho^2-v_r^2,
\end{aligned}
\label{eq:U12}
\end{equation}
where the last term comes from the rest-frame energy density, $E-\tfrac43\rho_rv_r^2$. In Fourier space $v_r^2\to-\tfrac{9}{16\hat\rho^2}(\bm\kappa_1\cdot\bm\kappa_2)\hat q(\bm\kappa_1)\hat q(\bm\kappa_2)$ and the advection $\bm v_r\cdot\nabla\delta\phi/aH\to-\tfrac{3}{4\hat\rho}(\bm\kappa_1\cdot\bm\kappa_2)\hat q(\bm\kappa_1)\delta\phi(\bm\kappa_2)$, and with $u^0=1+\tfrac12v_r^2$ and $h_n=V^{(n)}/H^2$ the sources are $\mathcal S_{\delta\phi}=0$ and
\begin{equation}
\begin{aligned}
\mathcal S_\pi={}&-\tfrac12h_3\,\delta\phi^2-3Q\big[v\,\mathcal U_2+\mathcal U_1\delta\pi+\tfrac12v\,v_r^2+\bm v_r\cdot\nabla\delta\phi\big],\\
\mathcal S_e={}&3Q\big[v^2\mathcal U_2+2v\,\mathcal U_1\delta\pi+\delta\pi^2+\tfrac12v^2v_r^2+v\,\bm v_r\cdot\nabla\delta\phi\big],\\
\mathcal S_q={}&\Big[\frac{3(\bm\kappa\cdot\bm\kappa_1)(\bm\kappa\cdot\bm\kappa_2)}{4\hat\rho\,\kappa^2}-\frac{\bm\kappa_1\cdot\bm\kappa_2}{4\hat\rho}\Big]\hat q(\bm\kappa_1)\hat q(\bm\kappa_2)\\
&-3Q\,\frac{\bm\kappa\cdot\bm\kappa_2}{\kappa^2}\big[v\,\mathcal U_1+\delta\pi\big](\bm\kappa_1)\,\delta\phi(\bm\kappa_2).
\end{aligned}
\label{eq:sources}
\end{equation}
The momentum source is the longitudinal part of the divergence of the anisotropic momentum flux $\tfrac43\rho_rv_r^iv_r^j$ and of the momentum transfer $\Upsilon u^\mu\partial_\mu\phi\,\partial_i\phi$. The periodic factor of the friction enters only through $\ell_1$ and $\ell_2$. Inside the window $\ell_1\simeq-(\delta_A/f)\sin\theta/(1+\delta_A\cos\theta)$, of order $24$, and $\tfrac12(\ell_2+\ell_1^2)=A''/2A\simeq-\tfrac12(\delta_A/f^2)\cos\theta/(1+\delta_A\cos\theta)$, of order $1.5\times10^3$. These are the periodic vertices. They make the expansion one in $\delta\phi/f$, and since $\mathcal P_{\delta\phi}^{1/2}/f$ is $0.01$--$0.02$ at crossing in the window, the periodic quadratic term is a percent-level correction to the periodic linear one.

The noise has the amplitude $\sqrt{2\Upsilon T\,S(\tilde z)}$, which depends on the local friction and temperature. Its relative change at first order is $m_1=\tfrac12[\ell_1\delta\phi+(1-s/4)\,e/\hat\rho]$, with $s=d\ln S/d\ln\tilde z$ at the momentum of the noise, and the noise also transfers energy and momentum to the radiation. The bilinear source is therefore $\mathcal N_c^{\,d}\hat\xi X_d$ with $\mathcal N_\pi^{\,d}=m_1^d$, $\mathcal N_e^{\,d}=-v\,m_1^d-\delta^d_\pi$ and $\mathcal N_q^{\,\delta\phi}=\bm\kappa\cdot\bm\kappa_2/\kappa^2$, where $m_1^d$ is the coefficient of $X_d$ in $m_1$ and $\bm\kappa_2$ the momentum of the field. A product of white noise and a field needs a convention, and we use It\^o's: the amplitude is evaluated at the start of each step, and the noise that drives a first-order leg correlates with the noise in $\mathcal S_\xi$ through the cross-variation $\sigma^2\hat n_a\mathcal N_c^{\,d}X_d\,dN$. The Stratonovich reading gives the same tree-level bispectrum. The noise on a first-order leg is additive, so converting between the two only adds the drift $\tfrac12\sum_{\bm p}\sigma^2(p)\mathcal N_c^{\,d}(\bm p,-\bm p)\hat n_d$, which sources the homogeneous mode and renormalizes the background. A physical noise of finite correlation time, whose white limit is Stratonovich by the Wong--Zakai theorem, therefore leads to the same result.

The tree-level moment $T_{abc}=\langle X_a(\bm k_1)X_b(\bm k_2)X^{(2)}_c(\bm k_3)\rangle'$ then obeys
\begin{equation}
\begin{aligned}
\frac{dT_{abc}}{dN}={}&\mathsf A_{aa'}(k_1)T_{a'bc}+\mathsf A_{bb'}(k_2)T_{ab'c}+\mathsf A_{cc'}(k_3)T_{abc'}\\
&+2V_c^{\,de}(\bm k_1,\bm k_2)\,\Sigma_{ad}(k_1)\Sigma_{be}(k_2)\\
&+\sigma^2(k_1)\,\hat n_a\mathcal N_c^{\;d}(\bm k_1,\bm k_2)\Sigma_{bd}(k_2)+(1\leftrightarrow2),
\end{aligned}
\label{eq:moments}
\end{equation}
with $V$ the symmetrized quadratic vertices of Eq.~\eqref{eq:sources} and $\Sigma$ the first-order covariances, and likewise for the other two choices of the second-order leg. The moments are driven by the same covariances that give the power spectrum and are integrated with them, with no further stochastic averaging, by an implicit Radau method at relative tolerance $10^{-7}$ for the moments and $10^{-8}$ for the covariances, from the time when the largest wavenumber has $z=100$; starting at $z=200$ changes $f_{\rm NL}$ by $6\times10^{-7}$ for an equilateral triangle and by $2\times10^{-5}$ for a squeezed one with $k_3/k_1=0.02$.

The curvature perturbation is the number of e-folds from a flat slice to the slice of uniform total density. Its first-order part is $\zeta_1=o\cdot X$ with $o=(h_1,v,1,0)/6\epsilon_H$, which is $\delta\rho_{\rm tot}/6\epsilon_HH^2$, and to second order
\begin{equation}
\zeta=o\cdot X+o\cdot X^{(2)}+\frac{\tfrac12h_2\delta\phi^2+\tfrac12\delta\pi^2}{6\epsilon_H}+\Big(\frac{\epsilon_N}{2\epsilon_H}-\epsilon_H\Big)\zeta_1^2,
\label{eq:dN}
\end{equation}
with $\epsilon_N=d\epsilon_H/dN$. The third term is the second-order energy density of the field, without its gradients outside the horizon, and the last combines the curvature of $\rho_{\rm tot}(N)$ with the change of the density perturbation on flat slices during the shift, $\delta\rho'=-\rho''_{\rm tot}\zeta_1$, which holds once $\zeta_1$ is conserved. The bispectrum thus has an evolution part, from the moments projected on $o$, and a local part, from the quadratic terms of Eq.~\eqref{eq:dN}. We use $f_{\rm NL}=\tfrac56B_\zeta/(P_1P_2+P_2P_3+P_3P_1)$ and read it off $3.5$ e-folds after the mode with the largest wavenumber crosses; between $3$ and $5$ e-folds it changes by at most $0.08$ in the window.

\subsection{Results}
\label{sec:bisresults}

\begin{figure*}[t]
\centering
\includegraphics[width=\textwidth]{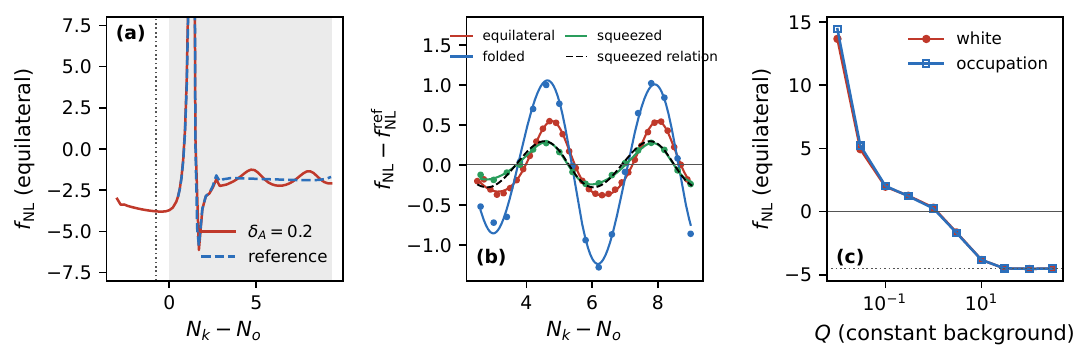}
\caption{The tree-level bispectrum. (a) The equilateral $f_{\rm NL}$ against the crossing time $N_k$ of $k_1$, measured from the opening of the window $N_o$, through the feature (dotted line: its peak) and the window (grey), with the periodic factor of the window on (red) and off (blue dashed). Near the spectral minimum between the two, $f_{\rm NL}=\tfrac56B/\sum P^2$ leaves the plotted range. (b) The oscillating part $f_{\rm NL}-f_{\rm NL}^{\rm ref}$ on the fitted modes for equilateral, folded ($k_2=k_3=k_1/2$) and squeezed ($k_1=k_2=10k_3$) triangles with the fitted template, and the oscillation that Eq.~\eqref{eq:fsq} gives from the computed spectrum (dashed). (c) The equilateral $f_{\rm NL}$ on constant backgrounds with $\Upsilon\propto T^3$ and $T/H=20$, for white and occupation-weighted noise; the dotted line marks $-4.5$.}
\label{fig:bispec}
\end{figure*}

Three checks come first. The quadratic vertices agree with a direct evaluation of the nonlinear equations on a periodic $16^3$ grid, with two plane waves of relative amplitude $10^{-5}$, to a relative $9\times10^{-5}$ for four random triangles in the window and at the peak. On constant backgrounds with $\Upsilon\propto T^3$, the equilateral $f_{\rm NL}$ is $+14$ at $Q=0.01$, changes sign between $Q=1$ and $3$, and tends to $-4.5$ at strong dissipation (Fig.~\ref{fig:bispec}). Reference~\cite{MirbabayiGruzinov2023} uses the same $\zeta=-H\delta\phi/\dot\phi$ but defines $f_{\rm NL}$ with the opposite sign, $f_{\rm NL}=-\tfrac56B/\sum PP$ (its Eq.~(64), with the squeezed limit $\tfrac5{12}(n_s-1)$), and it finds $+4.5$ in this limit (its Fig.~4), which is our $-4.5$. The growth at weak dissipation follows the trend of Refs.~\cite{BGMR2014NG,MirbabayiGruzinov2023}; the strong-regime $|f_{\rm NL}|\simeq10$ of Ref.~\cite{BGMR2014NG} is about twice ours, and that calculation includes a momentum-dependent suppression of the friction and a separate radiation noise, which we do not. Finally, the metric-free linear system reproduces the power of the full solver at read-off to within $0.3\%$ at the peak and in the first two thirds of the window, and to within $3\%$ for its last modes, whose read-off falls after the window closes, where $\epsilon_H$ rises.

Across the window the reference trajectory gives an equilateral $f_{\rm NL}$ between $-2.25$ and $-1.60$, close to the constant-background value at $Q\simeq3.3$, a folded one ($k_2=k_3=k_1/2$) near $5.6$ and a squeezed one ($k_1=k_2=10k_3$) near $-0.19$. With the periodic factor switched on, every configuration oscillates at the frequency of the comb (Fig.~\ref{fig:bispec}). We fit $f_{\rm NL}-f_{\rm NL}^{\rm ref}$ with the template of Eq.~\eqref{eq:fitmodel}, as an additive modulation at the lag of the power comb, so that its phase can be compared with $\psi_1$. The fit starts $2.5$ e-folds after the window opens, half an e-fold later than for the power, because the folded and squeezed triangles have sides that crossed earlier; starting at two e-folds lets the opening transient move the phases by up to half a radian, while across the later fit ranges they move by at most $0.04$~rad. The fundamental has amplitude $F_1=0.46$ in the equilateral configuration, $1.10$ folded and $0.25$ squeezed, with $\Delta\psi=2.13$, $1.89$ and $1.68$~rad. For every shape the coupling phase is evaluated at the crossing of the largest wavenumber, $k_1$; referring the folded triangle to its shorter sides instead would shift its $\Delta\psi$ by $\omega\ln2\simeq1.3$~rad.

The squeezed configuration tests the calculation. Applied to the computed spectrum with the same fit, Eq.~\eqref{eq:fsq} predicts $F_1=0.286$ with $\Delta\psi=1.59$~rad. For $k_3/k_1=0.1$ the tree-level moments give $F_1=0.254$ with $\Delta\psi=1.68$~rad, $11\%$ below the prediction in amplitude and $0.09$~rad away in phase. The difference is the finite ratio of the triangle: for $k_3/k_1=0.05$ and $0.02$ it falls to $4.0\%$ and $1.7\%$ in amplitude and to $0.033$ and $0.019$~rad in phase, computed and fitted in the same way. The relation does not use the modulated vertices, so the agreement checks the part of them that survives the soft limit, the response of the short modes to a long-wavelength shift of the background; the gradient vertices vanish in that limit and are checked by the direct evaluation above. Away from the squeezed limit $\Delta\psi$ moves away from $\pi/2$, by $0.32$~rad in the folded configuration and $0.56$~rad in the equilateral one. The oscillation is shared by all shapes, but its phase depends on the shape. With white noise the equilateral oscillation has amplitude $0.49$ and $\Delta\psi=2.14$~rad, on a reference value near $-1.82$, so these results do not hinge on the closure. In the bispectrum itself, $B\propto f_{\rm NL}P^2$, the equilateral oscillation of $f_{\rm NL}$, a quarter of its reference value, adds to the larger fractional oscillation $2\hat A_1\simeq0.68$ that the comb contributes through $P^2$.

At the black-hole peak the equilateral $f_{\rm NL}$ is $-3.8$, again close to the constant-background value at $Q\simeq9$, the folded one $9.8$ and the squeezed one $-0.69$. The diagnostics $|f_{\rm NL}|\Pz^{1/2}$ and $f_{\rm NL}^2\Pz$ are $0.31$ and $0.10$ there in the equilateral configuration, and at most $0.02$ and $5\times10^{-4}$ across the window. They are diagnostics, not corrections. The black-hole tail needs the statistics of the collapse variable~\cite{AtalGermani2019,Ferrante2023,Pi2024}, the induced peak needs the scalar four-point function~\cite{AdsheadLozanovWeiner2021}, and at the feature the tree-level calculation inherits the uncontrolled thermal onset. The window is locally much less nonlinear, but its small diagnostics do not bound loop integrals that reach back to the larger feature, so the comb and its gravitational waves remain linear-order results.

\section{Discussion and conclusions}
\label{sec:conclusions}

We have followed one idea, an inflaton whose dissipation is periodic in the field, through the background, the curvature spectrum, the induced gravitational waves and the bispectrum. The CMB scales cross with negligible dissipation and behave as in cold inflation. Their tilt is set by the plateau far above the dissipative region, so the benchmark value $n_s=0.964$ can be moved to the P-ACT-LB value $0.974$ by reshaping the plateau there, with the comb unchanged to $0.1\%$ and the black-hole feature retuned by $1.5\%$.

Where the dissipation stays moderately strong for several e-folds, the periodic friction writes a log-periodic comb on the spectrum, and the central result is how the coupled system responds to a friction that changes in time. We measured that response as a power-response kernel. Convolved with the explicit modulation, it reproduces the coupled comb, $\hat A_1\simeq0.34$ for the benchmark, to within about $7\%$ across five variations of the dissipation sector, and to within $3\%$ in four of them. The static response it measures lies $12$--$21\%$ below the estimate built from published growth-function fits. Our solver reproduces the slope of those fits when the noise is placed as in them, and with the energy-conserving, occupation-weighted noise used here its adiabatic estimate comes within $3\%$ of the static response. How much of the modulation survives is set by $\omega\Delta_N$, the modulation frequency times the width of freeze-out. In the tested window the modulus of the kernel follows the reduced filter $|\Gamma(3/2+i\omega\Delta_N)|/\Gamma(3/2)$ to within $3\%$, while its phase is set by a centroid about two e-folds before crossing. On the benchmark the comb changes by at most $4\%$ under four alternative noise closures, although the power itself moves by tens of percent.

The comb is what is specific to periodic dissipation, provided the periodic one-loop potential of the mediators stays small, as it does for a few supersymmetric mediator pairs (Sec.~\ref{sec:dissipation}). Log-periodic oscillations also come from resonant features in cold inflation~\cite{FlaugerDrifting2017} and survive into the induced gravitational waves in the same way~\cite{Fumagalli2021}, so an oscillation alone does not identify the mechanism. What would point to dissipation is a comb confined to a band of enhanced power, with an amplitude given by the static response of the friction times a filter set by $\omega\Delta_N$ and a frequency that drifts with the roll; we have not quantified how well these features discriminate against alternatives. In the induced gravitational waves the comb appears between $0.2$~Hz and $0.9$~kHz with $30$--$40\%$ contrast, and in an idealized forecast a BBO-class detector reaches a signal-to-noise ratio of 6--16 on it if the binary foreground is fully subtracted, and 2--8 if it is not. At tree level the bispectrum oscillates with the comb in every configuration, with a phase offset close to $\pi/2$ in the squeezed limit, where the consistency relation is reproduced, and of $2.1$~rad in the equilateral configuration. The equilateral $f_{\rm NL}$ is about $-1.8$ across the window and $-3.8$ at the black-hole peak.

A localized enhancement in the same sector gives, as a conditional Gaussian illustration, asteroid-mass black holes near $2\times10^{-13}\,\Msun$ with an order-unity dark-matter fraction for a tuned amplitude, and a peak $h^2\Ogw\simeq4.6\times10^{-10}$ near $19$~mHz. The other relations we find there are generic: the mass and the peak frequency are related by $M\propto f_{\rm pk}^{-2}$ for any radiation-era collapse, and the peak height scales as $(\Pz^{\rm pk})^2$ while the abundance is exponential.

The limitations are worth stating together. The dissipation coefficient is a phenomenological ansatz, and the microscopic sector that would give the cubic law, the mass hierarchy and loop control is not derived; in particular, a fermionic mediator on its own would give a periodic one-loop potential steeper than the plateau. With supersymmetry broken by the vacuum energy of the plateau, the remainder modulates the slope by $0.35$--$2.5\%$ per mediator pair, which keeps the oscillation it adds to the comb small only for a few pairs near the lower bound on $C_{\rm eff}$. The radiation is an ideal fluid, and the kernel has been measured only for $3<Q<5$ in one window; a finite memory of the noise moves the comb by at most $3.5\%$ for $(3H+\Upsilon)\tau_c\leq4$. At the onset of the feature the bath is heated faster than a weakly coupled plasma can relax, so the amplitude of the feature, and with it the abundance and the millihertz peak, are not under control; in the window the heating is slow, but a thermal bath there still needs a relaxation rate that we do not compute. The bispectrum is computed at tree level with the metric unperturbed, and its loop corrections, which could connect the feature to the window, are not computed. What the coupled calculation establishes is the comb and its size at linear order.

These limits suggest the next steps: a kinetic treatment of the bath through a fast onset, the kernel over a wider range of friction and with a viscous fluid, the statistics of the collapse variable built on the computed bispectrum, the loop corrections between the feature and the window, and, observationally, a decihertz detector able to subtract the binary foreground and search for the comb directly.

\begin{acknowledgments}
The work of M.R.G.\ is supported by the Science and Engineering Research Board (SERB), DST, Government of India, under Grant No.\ CRG/2022/004120 (Core Research Grant). M.R.G.\ thanks IUCAA, Pune, for hospitality during an associate visit where this work was initiated. 

We thank the anonymous referee for raising very valid points during the revision, which substantially changed the original manuscript; indeed, even the title has been changed accordingly. The referee report substantially improved the scientific quality of this work.
\end{acknowledgments}
\section*{Generative AI disclosure}
A generative AI, Claude (Anthropic) was used for debugging portions of the analysis
and plotting code, and as an aid in cross-checking bibliographic entries. All
models, derivations and physical conclusions are the author's own; the author produced every numerical result and figure with their own code.


\begin{thebibliography}{134}

\bibitem{BereraFang1995} A.~Berera and L.~Z.~Fang, Phys. Rev. Lett. \textbf{74}, 1912 (1995).
\bibitem{Berera1995} A.~Berera, Phys. Rev. Lett. \textbf{75}, 3218 (1995).
\bibitem{BereraGleiserRamos1998} A.~Berera, M.~Gleiser, and R.~O.~Ramos, Phys. Rev. D \textbf{58}, 123508 (1998), \href{https://arxiv.org/abs/hep-ph/9803394}{arXiv:hep-ph/9803394}.
\bibitem{BereraMossRamos2009} A.~Berera, I.~G.~Moss, and R.~O.~Ramos, Rept. Prog. Phys. \textbf{72}, 026901 (2009), \href{https://arxiv.org/abs/0808.1855}{arXiv:0808.1855}.
\bibitem{BasteroGilBerera2009} M.~Bastero-Gil and A.~Berera, Int. J. Mod. Phys. A \textbf{24}, 2207 (2009), \href{https://arxiv.org/abs/0902.0521}{arXiv:0902.0521}.
\bibitem{Berera1997} A.~Berera, Phys. Rev. D \textbf{55}, 3346 (1997), \href{https://arxiv.org/abs/hep-ph/9612239}{arXiv:hep-ph/9612239}.
\bibitem{YokoyamaLinde1999} J.~Yokoyama and A.~D.~Linde, Phys. Rev. D \textbf{60}, 083509 (1999), \href{https://arxiv.org/abs/hep-ph/9809409}{arXiv:hep-ph/9809409}.
\bibitem{BGRR2011diss} M.~Bastero-Gil, A.~Berera, and R.~O.~Ramos, JCAP \textbf{09}, 033 (2011), \href{https://arxiv.org/abs/1008.1929}{arXiv:1008.1929}.
\bibitem{BereraRamos2005} A.~Berera and R.~O.~Ramos, Phys. Rev. D \textbf{71}, 023513 (2005), \href{https://arxiv.org/abs/hep-ph/0406339}{arXiv:hep-ph/0406339}.
\bibitem{BereraKephart1999} A.~Berera and T.~W.~Kephart, Phys. Rev. Lett. \textbf{83}, 1084 (1999), \href{https://arxiv.org/abs/hep-ph/9904410}{arXiv:hep-ph/9904410}.
\bibitem{LaineProcacci2021} M.~Laine and S.~Procacci, JCAP \textbf{06}, 031 (2021), \href{https://arxiv.org/abs/2102.09913}{arXiv:2102.09913}.
\bibitem{WLI2016} M.~Bastero-Gil, A.~Berera, R.~O.~Ramos, and J.~G.~Rosa, Phys. Rev. Lett. \textbf{117}, 151301 (2016), \href{https://arxiv.org/abs/1604.08838}{arXiv:1604.08838}.
\bibitem{BGRR2013} M.~Bastero-Gil, A.~Berera, R.~O.~Ramos, and J.~G.~Rosa, JCAP \textbf{01}, 016 (2013), \href{https://arxiv.org/abs/1207.0445}{arXiv:1207.0445}.
\bibitem{ReliableEFT2021} M.~Bastero-Gil, A.~Berera, R.~O.~Ramos, and J.~G.~Rosa, Phys. Lett. B \textbf{813}, 136055 (2021), \href{https://arxiv.org/abs/1907.13410}{arXiv:1907.13410}.
\bibitem{BenettiRamos2017} M.~Benetti and R.~O.~Ramos, Phys. Rev. D \textbf{95}, 023517 (2017), \href{https://arxiv.org/abs/1610.08758}{arXiv:1610.08758}.
\bibitem{BasteroGilCMB2018} M.~Bastero-Gil, S.~Bhattacharya, K.~Dutta, and M.~R.~Gangopadhyay, JCAP \textbf{02}, 054 (2018), \href{https://arxiv.org/abs/1710.10008}{arXiv:1710.10008}.
\bibitem{Planck2018X} Y.~Akrami \textit{et al.} (Planck Collaboration), Astron. Astrophys. \textbf{641}, A10 (2020), \href{https://arxiv.org/abs/1807.06211}{arXiv:1807.06211}.
\bibitem{BereraMabillard2018} A.~Berera, J.~Mabillard, M.~Pieroni, and R.~O.~Ramos, JCAP \textbf{07}, 021 (2018), \href{https://arxiv.org/abs/1803.04982}{arXiv:1803.04982}.
\bibitem{GangopadhyayWQI2021} M.~R.~Gangopadhyay, S.~Myrzakul, M.~Sami, and M.~K.~Sharma, Phys. Rev. D \textbf{103}, 043505 (2021), \href{https://arxiv.org/abs/2011.09155}{arXiv:2011.09155}.
\bibitem{BasakWQI2022} S.~Basak, S.~Bhattacharya, M.~R.~Gangopadhyay, N.~Jaman, R.~Rangarajan, and M.~Sami, JCAP \textbf{03}, 063 (2022), \href{https://arxiv.org/abs/2110.00607}{arXiv:2110.00607}.
\bibitem{FreeseFriemanOlinto1990} K.~Freese, J.~A.~Frieman, and A.~V.~Olinto, Phys. Rev. Lett. \textbf{65}, 3233 (1990).
\bibitem{SilversteinWestphal2008} E.~Silverstein and A.~Westphal, Phys. Rev. D \textbf{78}, 106003 (2008), \href{https://arxiv.org/abs/0803.3085}{arXiv:0803.3085}.
\bibitem{McAllister2010} L.~McAllister, E.~Silverstein, and A.~Westphal, Phys. Rev. D \textbf{82}, 046003 (2010), \href{https://arxiv.org/abs/0808.0706}{arXiv:0808.0706}.
\bibitem{FlaugerOscillations2010} R.~Flauger, L.~McAllister, E.~Pajer, A.~Westphal, and G.~Xu, JCAP \textbf{06}, 009 (2010), \href{https://arxiv.org/abs/0907.2916}{arXiv:0907.2916}.
\bibitem{FlaugerPajer2011} R.~Flauger and E.~Pajer, JCAP \textbf{01}, 017 (2011), \href{https://arxiv.org/abs/1002.0833}{arXiv:1002.0833}.
\bibitem{Hannestad2010} S.~Hannestad, T.~Haugbolle, P.~R.~Jarnhus, and M.~S.~Sloth, JCAP \textbf{06}, 001 (2010), \href{https://arxiv.org/abs/0912.3527}{arXiv:0912.3527}.
\bibitem{FlaugerDrifting2017} R.~Flauger, L.~McAllister, E.~Silverstein, and A.~Westphal, JCAP \textbf{10}, 055 (2017), \href{https://arxiv.org/abs/1412.1814}{arXiv:1412.1814}.
\bibitem{Behbahani2012} S.~R.~Behbahani, A.~Dymarsky, M.~Mirbabayi, and L.~Senatore, JCAP \textbf{12}, 036 (2012), \href{https://arxiv.org/abs/1111.3373}{arXiv:1111.3373}.
\bibitem{ChenEastherLim2008} X.~Chen, R.~Easther, and E.~A.~Lim, JCAP \textbf{04}, 010 (2008), \href{https://arxiv.org/abs/0801.3295}{arXiv:0801.3295}.
\bibitem{ChenReview2010} X.~Chen, Adv. Astron. \textbf{2010}, 638979 (2010), \href{https://arxiv.org/abs/1002.1416}{arXiv:1002.1416}.
\bibitem{KobayashiTakahashi2011} T.~Kobayashi and F.~Takahashi, JCAP \textbf{01}, 026 (2011), \href{https://arxiv.org/abs/1011.3988}{arXiv:1011.3988}.
\bibitem{Achucarro2011} A.~Ach\'ucarro, J.-O.~Gong, S.~Hardeman, G.~A.~Palma, and S.~P.~Patil, JCAP \textbf{01}, 030 (2011), \href{https://arxiv.org/abs/1010.3693}{arXiv:1010.3693}.
\bibitem{ChenFeatures2012} X.~Chen, JCAP \textbf{01}, 038 (2012), \href{https://arxiv.org/abs/1104.1323}{arXiv:1104.1323}.
\bibitem{PajerPeloso2013} E.~Pajer and M.~Peloso, Class. Quant. Grav. \textbf{30}, 214002 (2013), \href{https://arxiv.org/abs/1305.3557}{arXiv:1305.3557}.
\bibitem{Maldacena2003} J.~M.~Maldacena, JHEP \textbf{05}, 013 (2003), \href{https://arxiv.org/abs/astro-ph/0210603}{arXiv:astro-ph/0210603}.
\bibitem{AcquavivaBartolo2003} V.~Acquaviva, N.~Bartolo, S.~Matarrese, and A.~Riotto, Nucl. Phys. B \textbf{667}, 119 (2003), \href{https://arxiv.org/abs/astro-ph/0209156}{arXiv:astro-ph/0209156}.
\bibitem{BabichCreminelli2004} D.~Babich, P.~Creminelli, and M.~Zaldarriaga, JCAP \textbf{08}, 009 (2004), \href{https://arxiv.org/abs/astro-ph/0405356}{arXiv:astro-ph/0405356}.
\bibitem{ChenHuangKachruShiu2007} X.~Chen, M.-x.~Huang, S.~Kachru, and G.~Shiu, JCAP \textbf{01}, 002 (2007), \href{https://arxiv.org/abs/hep-th/0605045}{arXiv:hep-th/0605045}.
\bibitem{WarmNatural2023} G.~Montefalcone, V.~Aragam, L.~Visinelli, and K.~Freese, JCAP \textbf{03}, 002 (2023), \href{https://arxiv.org/abs/2212.04482}{arXiv:2212.04482}.
\bibitem{FerrazRosa2023} P.~B.~Ferraz and J.~G.~Rosa, JHEP \textbf{12}, 176 (2023), \href{https://arxiv.org/abs/2308.00564}{arXiv:2308.00564}.
\bibitem{GorgulhoRosa2025} D.~S.~Gorgulho and J.~G.~Rosa, JHEP \textbf{04}, 124 (2026), \href{https://arxiv.org/abs/2507.13156}{arXiv:2507.13156}.
\bibitem{Arya2020} R.~Arya, JCAP \textbf{09}, 042 (2020), \href{https://arxiv.org/abs/1910.05238}{arXiv:1910.05238}.
\bibitem{BasteroGilSubias2021} M.~Bastero-Gil and M.~Sub\'{\i}as D\'{\i}az-Blanco, JCAP \textbf{12}, 052 (2021), \href{https://arxiv.org/abs/2105.08045}{arXiv:2105.08045}.
\bibitem{CorreaWNI2022} M.~Correa, M.~R.~Gangopadhyay, N.~Jaman, and G.~J.~Mathews, Phys. Lett. B \textbf{835}, 137510 (2022), \href{https://arxiv.org/abs/2207.10394}{arXiv:2207.10394}.
\bibitem{Correa2024GW} M.~Correa, M.~R.~Gangopadhyay, N.~Jaman, and G.~J.~Mathews, Phys. Rev. D \textbf{109}, 063539 (2024), \href{https://arxiv.org/abs/2306.09641}{arXiv:2306.09641}.
\bibitem{AryaJainMishra2024} R.~Arya, R.~K.~Jain, and A.~K.~Mishra, JCAP \textbf{02}, 034 (2024), \href{https://arxiv.org/abs/2302.08940}{arXiv:2302.08940}.
\bibitem{FerrazMansoRosa2026} P.~B.~Ferraz, A.~Torres~Manso, and J.~G.~Rosa, \href{https://arxiv.org/abs/2609.19270}{arXiv:2609.19270}.
\bibitem{ItoRamos2025} A.~Ito and R.~O.~Ramos, JCAP \textbf{08}, 076 (2025), \href{https://arxiv.org/abs/2504.15606}{arXiv:2504.15606}.
\bibitem{Ballesteros2022} G.~Ballesteros, M.~A.~G.~Garc\'ia, A.~P\'erez Rodr\'iguez, M.~Pierre, and J.~Rey, JCAP \textbf{12}, 006 (2022), \href{https://arxiv.org/abs/2208.14978}{arXiv:2208.14978}.
\bibitem{Fumagalli2021} J.~Fumagalli, S.~Renaux-Petel, and L.~T.~Witkowski, JCAP \textbf{08}, 059 (2021), \href{https://arxiv.org/abs/2105.06481}{arXiv:2105.06481}.
\bibitem{Witkowski2022} L.~T.~Witkowski, G.~Dom\`enech, J.~Fumagalli, and S.~Renaux-Petel, JCAP \textbf{05}, 028 (2022), \href{https://arxiv.org/abs/2110.09480}{arXiv:2110.09480}.
\bibitem{CalcagniKuroyanagi2024} G.~Calcagni and S.~Kuroyanagi, Class. Quant. Grav. \textbf{41}, 015031 (2024), \href{https://arxiv.org/abs/2308.05904}{arXiv:2308.05904}.
\bibitem{Padilla2026} L.~E.~Padilla, T.~Harada, E.~Milligan, and D.~Mulryne, Phys. Rev. D \textbf{114}, L041307 (2026), \href{https://arxiv.org/abs/2604.21520}{arXiv:2604.21520}.
\bibitem{Padilla2026b} L.~E.~Padilla, T.~Harada, and H.~Iizuka, \href{https://arxiv.org/abs/2608.21834}{arXiv:2608.21834}.
\bibitem{CaiTongWangYan2018} Y.-F.~Cai, X.~Tong, D.-G.~Wang, and S.-F.~Yan, Phys. Rev. Lett. \textbf{121}, 081306 (2018), \href{https://arxiv.org/abs/1805.03639}{arXiv:1805.03639}.
\bibitem{Zhou2020} Z.~Zhou, J.~Jiang, Y.-F.~Cai, M.~Sasaki, and S.~Pi, \href{https://arxiv.org/abs/2010.03537}{arXiv:2010.03537}.
\bibitem{MinimalWarmInflation} K.~V.~Berghaus, P.~W.~Graham, and D.~E.~Kaplan, JCAP \textbf{03}, 034 (2020) [Erratum: JCAP \textbf{10}, E02 (2023)], \href{https://arxiv.org/abs/1910.07525}{arXiv:1910.07525}.
\bibitem{WarmBaryogenesis2012} M.~Bastero-Gil, A.~Berera, R.~O.~Ramos, and J.~G.~Rosa, Phys. Lett. B \textbf{712}, 425 (2012), \href{https://arxiv.org/abs/1110.3971}{arXiv:1110.3971}.
\bibitem{RosaVentura2019} J.~G.~Rosa and L.~B.~Ventura, Phys. Rev. Lett. \textbf{122}, 161301 (2019), \href{https://arxiv.org/abs/1811.05493}{arXiv:1811.05493}.
\bibitem{Zell2025} S.~Zell, Phys. Rev. D \textbf{112}, L081307 (2025), \href{https://arxiv.org/abs/2408.07746}{arXiv:2408.07746} [hep-ph].
\bibitem{KimNillesPeloso2005} J.~E.~Kim, H.~P.~Nilles, and M.~Peloso, JCAP \textbf{01}, 005 (2005), \href{https://arxiv.org/abs/hep-ph/0409138}{arXiv:hep-ph/0409138}.
\bibitem{ColemanWeinberg1973} S.~R.~Coleman and E.~J.~Weinberg, Phys. Rev. D \textbf{7}, 1888 (1973).
\bibitem{BK18} P.~A.~R.~Ade \textit{et al.} (BICEP/Keck Collaboration), Phys. Rev. Lett. \textbf{127}, 151301 (2021), \href{https://arxiv.org/abs/2110.00483}{arXiv:2110.00483}.
\bibitem{ACTDR6} T.~Louis \emph{et al.} (ACT Collaboration), JCAP \textbf{11}, 062 (2025), \href{https://arxiv.org/abs/2503.14452}{arXiv:2503.14452}.
\bibitem{BGRR2018Boltzmann} M.~Bastero-Gil, A.~Berera, R.~O.~Ramos, and J.~G.~Rosa, JHEP \textbf{02}, 063 (2018), \href{https://arxiv.org/abs/1711.09023}{arXiv:1711.09023}.
\bibitem{RamosdaSilva2013} R.~O.~Ramos and L.~A.~da Silva, JCAP \textbf{03}, 032 (2013), \href{https://arxiv.org/abs/1302.3544}{arXiv:1302.3544}.
\bibitem{BGRR2014fluid} M.~Bastero-Gil, A.~Berera, I.~G.~Moss, and R.~O.~Ramos, JCAP \textbf{05}, 004 (2014), \href{https://arxiv.org/abs/1401.1149}{arXiv:1401.1149}.
\bibitem{ShearViscous2011} M.~Bastero-Gil, A.~Berera, and R.~O.~Ramos, JCAP \textbf{07}, 030 (2011), \href{https://arxiv.org/abs/1106.0701}{arXiv:1106.0701}.
\bibitem{LaineProcacciRogelj2025} M.~Laine, S.~Procacci, and A.~Rogelj, JCAP \textbf{12}, 058 (2025), \href{https://arxiv.org/abs/2507.12849}{arXiv:2507.12849}.
\bibitem{Montefalcone2024} G.~Montefalcone, V.~Aragam, L.~Visinelli, and K.~Freese, JCAP \textbf{01}, 032 (2024), \href{https://arxiv.org/abs/2306.16190}{arXiv:2306.16190}.
\bibitem{GrahamMoss2009} C.~Graham and I.~G.~Moss, JCAP \textbf{07}, 013 (2009), \href{https://arxiv.org/abs/0905.3500}{arXiv:0905.3500}.
\bibitem{Visinelli2015} L.~Visinelli, JCAP \textbf{01}, 005 (2015), \href{https://arxiv.org/abs/1410.1187}{arXiv:1410.1187}.
\bibitem{FlucDiss2018} M.~Bastero-Gil, A.~Berera, R.~Brandenberger, I.~G.~Moss, R.~O.~Ramos, and J.~G.~Rosa, JCAP \textbf{01}, 002 (2018), \href{https://arxiv.org/abs/1612.04726}{arXiv:1612.04726}.
\bibitem{GangopadhyayPBH2022} M.~R.~Gangopadhyay, J.~C.~Jain, D.~Sharma, and Yogesh, Eur. Phys. J. C \textbf{82}, 849 (2022), \href{https://arxiv.org/abs/2108.13839}{arXiv:2108.13839}.
\bibitem{ChoudhuryNoGo2023} S.~Choudhury, M.~R.~Gangopadhyay, and M.~Sami, Eur. Phys. J. C \textbf{84}, 884 (2024), \href{https://arxiv.org/abs/2301.10000}{arXiv:2301.10000}.
\bibitem{GangopadhyayKumar2026} M.~R.~Gangopadhyay and N.~Kumar, Phys. Dark Univ. \textbf{52}, 102355 (2026), \href{https://arxiv.org/abs/2603.11629}{arXiv:2603.11629}.
\bibitem{Stewart2002} E.~D.~Stewart, Phys. Rev. D \textbf{65}, 103508 (2002), \href{https://arxiv.org/abs/astro-ph/0110322}{arXiv:astro-ph/0110322}.
\bibitem{DvorkinHu2010} C.~Dvorkin and W.~Hu, Phys. Rev. D \textbf{81}, 023518 (2010), \href{https://arxiv.org/abs/0910.2237}{arXiv:0910.2237}.
\bibitem{Musco2019} I.~Musco, Phys. Rev. D \textbf{100}, 123524 (2019), \href{https://arxiv.org/abs/1809.02127}{arXiv:1809.02127}.
\bibitem{YoungByrnesSasaki2014} S.~Young, C.~T.~Byrnes, and M.~Sasaki, JCAP \textbf{07}, 045 (2014), \href{https://arxiv.org/abs/1405.7023}{arXiv:1405.7023}.
\bibitem{HaradaYooKohri2013} T.~Harada, C.-M.~Yoo, and K.~Kohri, Phys. Rev. D \textbf{88}, 084051 (2013) [Erratum: Phys. Rev. D \textbf{89}, 029903 (2014)], \href{https://arxiv.org/abs/1309.4201}{arXiv:1309.4201}.
\bibitem{PressSchechter1974} W.~H.~Press and P.~Schechter, Astrophys. J. \textbf{187}, 425 (1974).
\bibitem{CarrHawking1974} B.~J.~Carr and S.~W.~Hawking, Mon. Not. R. Astron. Soc. \textbf{168}, 399 (1974).
\bibitem{Carr1975} B.~J.~Carr, Astrophys. J. \textbf{201}, 1 (1975).
\bibitem{Musco2021} I.~Musco, V.~De~Luca, G.~Franciolini, and A.~Riotto, Phys. Rev. D \textbf{103}, 063538 (2021), \href{https://arxiv.org/abs/2011.03014}{arXiv:2011.03014}.
\bibitem{Ferrante2023} G.~Ferrante, G.~Franciolini, A.~J.~Iovino, and A.~Urbano, Phys. Rev. D \textbf{107}, 043520 (2023), \href{https://arxiv.org/abs/2211.01728}{arXiv:2211.01728}.
\bibitem{Choptuik1993} M.~W.~Choptuik, Phys. Rev. Lett. \textbf{70}, 9 (1993).
\bibitem{NiemeyerJedamzik1999} J.~C.~Niemeyer and K.~Jedamzik, Phys. Rev. D \textbf{59}, 124013 (1999), \href{https://arxiv.org/abs/astro-ph/9901292}{arXiv:astro-ph/9901292}.
\bibitem{ShibataSasaki1999} M.~Shibata and M.~Sasaki, Phys. Rev. D \textbf{60}, 084002 (1999), \href{https://arxiv.org/abs/gr-qc/9905064}{arXiv:gr-qc/9905064}.
\bibitem{EvansColeman1994} C.~R.~Evans and J.~S.~Coleman, Phys. Rev. Lett. \textbf{72}, 1782 (1994), \href{https://arxiv.org/abs/gr-qc/9402041}{arXiv:gr-qc/9402041}.
\bibitem{MuscoMiller2013} I.~Musco and J.~C.~Miller, Class. Quant. Grav. \textbf{30}, 145009 (2013), \href{https://arxiv.org/abs/1201.2379}{arXiv:1201.2379}.
\bibitem{GreenKavanagh2021} A.~M.~Green and B.~J.~Kavanagh, J. Phys. G \textbf{48}, 043001 (2021), \href{https://arxiv.org/abs/2007.10722}{arXiv:2007.10722}.
\bibitem{CarrKohriSendouda2021} B.~Carr, K.~Kohri, Y.~Sendouda, and J.~Yokoyama, Rept. Prog. Phys. \textbf{84}, 116902 (2021), \href{https://arxiv.org/abs/2002.12778}{arXiv:2002.12778}.
\bibitem{Niikura2019HSC} H.~Niikura \textit{et al.}, Nat. Astron. \textbf{3}, 524 (2019), \href{https://arxiv.org/abs/1701.02151}{arXiv:1701.02151}.
\bibitem{Niikura2019OGLE} H.~Niikura, M.~Takada, S.~Yokoyama, T.~Sumi, and S.~Masaki, Phys. Rev. D \textbf{99}, 083503 (2019), \href{https://arxiv.org/abs/1901.07120}{arXiv:1901.07120}.
\bibitem{CarrKuhnel2020} B.~Carr and F.~K\"uhnel, Ann. Rev. Nucl. Part. Sci. \textbf{70}, 355 (2020), \href{https://arxiv.org/abs/2006.02838}{arXiv:2006.02838}.
\bibitem{CarrKuhnelVisinelli2021} B.~Carr, F.~K\"uhnel, and L.~Visinelli, Mon. Not. R. Astron. Soc. \textbf{501}, 2029 (2021), \href{https://arxiv.org/abs/2008.08077}{arXiv:2008.08077}.
\bibitem{SasakiSuyama2018} M.~Sasaki, T.~Suyama, T.~Tanaka, and S.~Yokoyama, Class. Quant. Grav. \textbf{35}, 063001 (2018), \href{https://arxiv.org/abs/1801.05235}{arXiv:1801.05235}.
\bibitem{CarrKuhnelSandstad2016} B.~Carr, F.~K\"uhnel, and M.~Sandstad, Phys. Rev. D \textbf{94}, 083504 (2016), \href{https://arxiv.org/abs/1607.06077}{arXiv:1607.06077}.
\bibitem{YooHaradaGarriga2018} C.-M.~Yoo, T.~Harada, J.~Garriga, and K.~Kohri, PTEP \textbf{2018}, 123E01 (2018); \textbf{2024}, 049202(E) (2024), \href{https://arxiv.org/abs/1805.03946}{arXiv:1805.03946}.
\bibitem{AtalGermani2019} V.~Atal and C.~Germani, Phys. Dark Univ. \textbf{24}, 100275 (2019), \href{https://arxiv.org/abs/1811.07857}{arXiv:1811.07857}.
\bibitem{Pi2024} S.~Pi, in \emph{Primordial Black Holes}, edited by C.~Byrnes, G.~Franciolini, T.~Harada, P.~Pani, and M.~Sasaki, Springer Series in Astrophysics and Cosmology (Springer, Singapore, 2025), pp.~155--200, \href{https://arxiv.org/abs/2404.06151}{arXiv:2404.06151}.
\bibitem{AnandaClarksonWands2007} K.~N.~Ananda, C.~Clarkson, and D.~Wands, Phys. Rev. D \textbf{75}, 123518 (2007), \href{https://arxiv.org/abs/gr-qc/0612013}{arXiv:gr-qc/0612013}.
\bibitem{BaumannSteinhardt2007} D.~Baumann, P.~J.~Steinhardt, K.~Takahashi, and K.~Ichiki, Phys. Rev. D \textbf{76}, 084019 (2007), \href{https://arxiv.org/abs/hep-th/0703290}{arXiv:hep-th/0703290}.
\bibitem{SaitoYokoyama2009} R.~Saito and J.~Yokoyama, Phys. Rev. Lett. \textbf{102}, 161101 (2009); \textbf{107}, 069901(E) (2011), \href{https://arxiv.org/abs/0812.4339}{arXiv:0812.4339}.
\bibitem{BartoloDeLuca2019} N.~Bartolo, V.~De Luca, G.~Franciolini, M.~Peloso, D.~Racco, and A.~Riotto, Phys. Rev. D \textbf{99}, 103521 (2019), \href{https://arxiv.org/abs/1810.12224}{arXiv:1810.12224}.
\bibitem{CapriniFigueroa2018} C.~Caprini and D.~G.~Figueroa, Class. Quant. Grav. \textbf{35}, 163001 (2018), \href{https://arxiv.org/abs/1801.04268}{arXiv:1801.04268}.
\bibitem{Domenech2021} G.~Dom\`enech, Universe \textbf{7}, 398 (2021), \href{https://arxiv.org/abs/2109.01398}{arXiv:2109.01398}.
\bibitem{KohriTerada2018} K.~Kohri and T.~Terada, Phys. Rev. D \textbf{97}, 123532 (2018), \href{https://arxiv.org/abs/1804.08577}{arXiv:1804.08577}.
\bibitem{EspinosaRaccoRiotto2018} J.~R.~Espinosa, D.~Racco, and A.~Riotto, JCAP \textbf{09}, 012 (2018), \href{https://arxiv.org/abs/1804.07732}{arXiv:1804.07732}.
\bibitem{LISA2017} P.~Amaro-Seoane \textit{et al.} (LISA Collaboration), \href{https://arxiv.org/abs/1702.00786}{arXiv:1702.00786}.
\bibitem{LISA2024} M.~Colpi \textit{et al.} (LISA Collaboration), \href{https://arxiv.org/abs/2402.07571}{arXiv:2402.07571}.
\bibitem{ThraneRomano2013} E.~Thrane and J.~D.~Romano, Phys. Rev. D \textbf{88}, 124032 (2013), \href{https://arxiv.org/abs/1310.5300}{arXiv:1310.5300}.
\bibitem{RobsonCornishLiu2019} T.~Robson, N.~J.~Cornish, and C.~Liu, Class. Quant. Grav. \textbf{36}, 105011 (2019), \href{https://arxiv.org/abs/1803.01944}{arXiv:1803.01944}.
\bibitem{AdsheadLozanovWeiner2021} P.~Adshead, K.~D.~Lozanov, and Z.~J.~Weiner, JCAP \textbf{10}, 080 (2021), \href{https://arxiv.org/abs/2105.01659}{arXiv:2105.01659}.
\bibitem{CaiPiSasaki2019} R.-G.~Cai, S.~Pi, and M.~Sasaki, Phys. Rev. Lett. \textbf{122}, 201101 (2019), \href{https://arxiv.org/abs/1810.11000}{arXiv:1810.11000}.
\bibitem{YuanChenHuang2020} C.~Yuan, Z.-C.~Chen, and Q.-G.~Huang, Phys. Rev. D \textbf{101}, 043019 (2020), \href{https://arxiv.org/abs/1910.09099}{arXiv:1910.09099}.
\bibitem{Kawamura2011} S.~Kawamura \textit{et al.}, Class. Quant. Grav. \textbf{28}, 094011 (2011).
\bibitem{YagiSeto2011} K.~Yagi and N.~Seto, Phys.\ Rev.\ D \textbf{83}, 044011 (2011); \textbf{95}, 109901(E) (2017), \href{https://arxiv.org/abs/1101.3940}{arXiv:1101.3940}.
\bibitem{NANOGrav2023} G.~Agazie \textit{et al.} (NANOGrav Collaboration), Astrophys. J. Lett. \textbf{951}, L8 (2023), \href{https://arxiv.org/abs/2306.16213}{arXiv:2306.16213}.
\bibitem{Punturo2010} M.~Punturo \textit{et al.}, Class. Quant. Grav. \textbf{27}, 194002 (2010).
\bibitem{Maggiore2020} M.~Maggiore \textit{et al.}, JCAP \textbf{03}, 050 (2020), \href{https://arxiv.org/abs/1912.02622}{arXiv:1912.02622}.
\bibitem{Yagi2013} K.~Yagi, Int. J. Mod. Phys. D \textbf{22}, 1341013 (2013), \href{https://arxiv.org/abs/1302.2388}{arXiv:1302.2388}.
\bibitem{Schmitz2020} K.~Schmitz, JHEP \textbf{01}, 097 (2021), \href{https://arxiv.org/abs/2002.04615}{arXiv:2002.04615} [hep-ph].
\bibitem{MossXiong2007} I.~G.~Moss and C.~Xiong, JCAP \textbf{04}, 007 (2007), \href{https://arxiv.org/abs/astro-ph/0701302}{arXiv:astro-ph/0701302}.
\bibitem{MossYeomans2011} I.~G.~Moss and T.~Yeomans, JCAP \textbf{08}, 009 (2011), \href{https://arxiv.org/abs/1102.2833}{arXiv:1102.2833}.
\bibitem{BGMR2014NG} M.~Bastero-Gil, A.~Berera, I.~G.~Moss, and R.~O.~Ramos, JCAP \textbf{12}, 008 (2014), \href{https://arxiv.org/abs/1408.4391}{arXiv:1408.4391}.
\bibitem{MirbabayiGruzinov2023} M.~Mirbabayi and A.~Gruzinov, JCAP \textbf{02}, 012 (2023), \href{https://arxiv.org/abs/2205.13227}{arXiv:2205.13227}.
\bibitem{Planck2018IX} Y.~Akrami \textit{et al.} (Planck Collaboration), Astron. Astrophys. \textbf{641}, A9 (2020), \href{https://arxiv.org/abs/1905.05697}{arXiv:1905.05697}.
\bibitem{CreminelliZaldarriaga2004} P.~Creminelli and M.~Zaldarriaga, JCAP \textbf{10}, 006 (2004), \href{https://arxiv.org/abs/astro-ph/0407059}{arXiv:astro-ph/0407059}.
\bibitem{TanakaUrakawa2011} T.~Tanaka and Y.~Urakawa, JCAP \textbf{05}, 014 (2011), \href{https://arxiv.org/abs/1103.1251}{arXiv:1103.1251}.
\bibitem{PajerSchmidtZaldarriaga2013} E.~Pajer, F.~Schmidt, and M.~Zaldarriaga, Phys. Rev. D \textbf{88}, 083502 (2013), \href{https://arxiv.org/abs/1305.0824}{arXiv:1305.0824}.
\bibitem{BehbahaniGreen2012} S.~R.~Behbahani and D.~Green, JCAP \textbf{11}, 056 (2012), \href{https://arxiv.org/abs/1207.2779}{arXiv:1207.2779}.
\bibitem{AdsheadDvorkin2012} P.~Adshead, C.~Dvorkin, W.~Hu, and E.~A.~Lim, Phys. Rev. D \textbf{85}, 023531 (2012), \href{https://arxiv.org/abs/1110.3050}{arXiv:1110.3050}.

\end{thebibliography}
\end{document}